\documentclass{aa}  
\usepackage{natbib}
\usepackage[colorlinks=true,citecolor=blue]{hyperref}
\usepackage{comment}
\usepackage{amsmath}
\usepackage{graphicx}
\usepackage{rotating}
\usepackage{float}
\usepackage{threeparttable}
\usepackage{tablefootnote}

\defcitealias{Bilicki2018}{B18}
\defcitealias{Bilicki2021}{B21}

\newcommand{\orcid}[1]{\href{https://orcid.org/#1}{\includegraphics[scale=0.04]{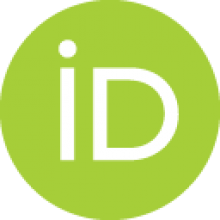}\hspace{.2mm}}}
\usepackage[normalem]{ulem}

\usepackage{graphicx}
\usepackage{txfonts}
\begin{document} 

\title{Enhancing photometric redshift catalogs through color-space analysis: Application to KiDS-bright galaxies}
\titlerunning{Color-Space Analysis of KiDS-Bright galaxies} 
\authorrunning{Jalan, Bilicki, et al.} 
\author{Priyanka Jalan                   \orcid{0000-0002-0524-5328} \inst{1} \and 
Maciej Bilicki           \orcid{0000-0002-3910-5809} \inst{1} \and 
Wojciech A. Hellwing             \orcid{0000-0003-4634-4442} \inst{1} \and 
Angus H. Wright          \orcid{0000-0001-7363-7932} \inst{2} \and\\ 
Andrej Dvornik           \orcid{0000-0001-7415-1977} \inst{2} \and 
Christos Georgiou \inst{3} \and
Catherine Heymans \inst{4,2} \and 
Hendrik Hildebrandt     \orcid{0000-0002-9814-3338} \inst{2} \and 
Shahab Joudaki \inst{5,6} \and 
Konrad Kuijken \orcid{0000-0002-3827-0175} \inst{7}  \and 
Constance Mahony \inst{8,2} \and 
Szymon Jan Nakoneczny \inst{9}  \orcid{0000-0003-2130-7143} \and 
Mario Radovich  \inst{10}       \orcid{0000-0002-3585-866X} \and 
Jan Luca van den Busch \inst{2} \and
Ziang Yan \inst{2} \and
Mijin Yoon \inst{6}}

\institute{Center for Theoretical Physics, Polish Academy of Sciences, Al. Lotników 32/46, 02-668,  Warsaw, Poland \\ \email{priyajalan14@gmail.com, pjalan@cft.edu.pl}
\and Ruhr University Bochum, Faculty of Physics and Astronomy, Astronomical Institute (AIRUB), German Centre for Cosmological Lensing, 44780 Bochum, Germany  
\and Institute for Theoretical Physics, Utrecht University, Princetonplein 5, 3584 CC, Utrecht, The Netherlands
\and Institute for Astronomy, University of Edinburgh, Royal Observatory, Blackford Hill, Edinburgh, EH9 3HJ, UK 
\and Centro de Investigaciones Energ\'{e} ticas, Medioambientales y Tecnol\'{o} gicas (CIEMAT), Av. Complutense 40, E-28040 Madrid, Spain 
\and Institute of Cosmology \& Gravitation, Dennis Sciama Building, University of Portsmouth, Portsmouth, PO1 3FX, UK 
\and Leiden Observatory, Leiden University, PO Box 9513, 2300 RA Leiden, the Netherlands 
\and Donostia International Physics Center, Manuel Lardizabal Ibilbidea, 4, 20018 Donostia, Gipuzkoa, Spain 
\and Division of Physics, Mathematics and Astronomy, California Institute of Technology, 1200 E California Blvd, Pasadena, CA 91125 
\and INAF - Osservatorio Astronomico di Padova, via dell'Osservatorio 5, 35122 Padova, Italy}

  \date{Received 30 September 2024 / Accepted 31 October 2024}

 
  \abstract
  {}
   {We present a method  for refining photometric redshift galaxy catalogs based on a comparison of their color-space matching with overlapping spectroscopic calibration data. We focus on cases where photometric redshifts (photo-$z$) are estimated empirically. Identifying galaxies that are poorly represented in spectroscopic data is crucial, as their photo-$z$ may be unreliable due to extrapolation beyond the training sample.}{Our approach uses a self-organizing map (SOM) to project a multidimensional parameter space of magnitudes and colors onto a 2D manifold, allowing us to analyze the resulting patterns as a function of various galaxy properties. Using SOM, we compared the  Kilo-Degree Survey's bright galaxy sample (KiDS-Bright), limited to $r<20$ mag, with various spectroscopic samples, including the Galaxy And Mass Assembly (GAMA).}{ Our analysis reveals that GAMA tends to underrepresent KiDS-Bright at its faintest ($r\gtrsim19.5$) and highest-redshift ($z\gtrsim0.4$) ranges; however, no strong trends are seen in terms of color or stellar mass. By incorporating additional spectroscopic data from the SDSS, 2dF, and early DESI, we identified SOM cells where the photo-$z$ values are estimated suboptimally. We derived a set of SOM-based criteria to refine the photometric sample and improve photo-$z$ statistics. For the KiDS-Bright sample, this improvement is modest, namely, it excludes the least represented 20\% of the sample reduces photo-$z$ scatter by less than 10\%.}{We conclude that GAMA, used for KiDS-Bright photo-$z$ training, is sufficiently representative for reliable redshift estimation across most of the color space. Future spectroscopic data from surveys such as DESI should be better suited for exploiting the full improvement potential of our method.}


   \maketitle




\section{Introduction}
\label{s1:intro}
Deep wide-angle imaging of the sky gives access to key probes of cosmology. Among these, weak gravitational lensing by the large-scale structure stands out as it allows us to study the distribution of both dark and luminous matter, as well as the overall properties of the Universe \citep[][]{Bartelmann2001, Hoekstra2008}. Current `Stage-III' imaging surveys such as the Dark Energy Survey \citep[DES,][]{DES2005}, Hyper Suprime-Cam Subaru Strategic Program \citep[HSC,][]{HSC2018}, and Kilo-Degree Survey \citep[KiDS,][]{Kuijken2019} have already covered thousands of square degrees and cataloged hundreds of millions of galaxies. The operational Euclid telescope \citep{EUCLID2012} and soon-to-be-launched Legacy Survey of Space and Time \citep{LSST2019} at Vera Rubin Observatory will mark the advent of the Stage-IV era, when most of the extragalactic sky will be covered, providing observations of billions of galaxies.

The most powerful cosmological signal from deep imaging surveys is the cosmic shear \citep[see][for a review]{Kilbinger2015}: the effect of coherent shape distortions of background galaxies (sources) due to the gravitational potentials of the large-scale matter distribution. Cosmic shear studies with these observations are most suitable for constraining two key cosmological parameters: the total non-relativistic matter density, $\Omega_\mathrm{m}$, and the amplitude of its fluctuations at 8 Mpc$/h$ scales,  where $\sigma_8$ is usually combined into $S_8 \equiv \sigma_8 \sqrt{\Omega_\mathrm{m}/0.3}$ \citep[e.g.,][]{Asgari2021, Amon2022DES, Li2023KiDS, Li2023HSC, Abbott2023DESKIDS}.

The constraints on $\Omega_\mathrm{m}$ and $\sigma_8$ from cosmic shear are degenerate, so extra measurements are needed to separate them. This is in particular possible thanks to the so-called multiprobe approach, where other observables are also used. Among these, the 3$\times$2-point approach has been successfully applied \citep[e.g.,][]{Heymans2021, Abbott2022DES, Miyatake2023HSC}. It combines cosmic shear, galaxy-galaxy lensing (GGL; shape distortions of background sources by foreground lenses), and clustering of the lenses.

The two main ways of selecting foreground galaxies for multiprobe analyses are either to use spectroscopic data that overlap (at least partly) with the imaging survey or by extracting these galaxies from the photometric sample itself. This former method presents several advantages, such as access to high-precision spectroscopic redshift estimates (spec-$z$) and the ability to measure redshift-space distortions and baryon acoustic oscillations from 3D clustering. This allows us to strengthen the cosmological constraints and break degeneracies between cosmological parameters. However, until now, the overlap of imaging datasets with appropriate wide-angle spectroscopic surveys for cosmic shear analysis and the number densities of the latter have been far from ideal. This leads, for instance, to the GGL signal being sub-dominant with respect to cosmic shear and clustering \citep[e.g.,][]{Heymans2021}, effectively reducing the analysis to 2$\times$2-point if the spectroscopic foreground is used.

In this context, selecting the foreground galaxies directly from the imaging survey is often beneficial, as it offers full-survey coverage, dense sampling, and control over the sample selection function. This comes at the cost, however, of redshift estimate precision: photometric redshift estimates (photo-$z$), derived directly from multiband imaging datasets, are typically of considerably lower quality than spectroscopic estimates.  This, in turn, allows us to measure only two-dimensional (projected) clustering in redshift bins. In addition, foreground galaxy selection from photometric data risks propagating observational systematic selection effects from the imaging into the lens sample, which can affect the clustering signal and redshift estimates. Furthermore, such systematics are generally more difficult to compensate for in photometric than in spectroscopic redshift surveys, due to a lack of precise information on the radial dimension. When using photometric galaxies for clustering measurements or as lenses in GGL, the quality requirements for their photo-$z$ are more stringent than those for source galaxies. Source galaxies are typically grouped in relatively broad redshift bins, making it crucial to constrain the population redshift distributions accurately. Conversely, for lenses, it is important to have precise knowledge of individual photo-$z$ and their uncertainties.

Several approaches exist for selecting photometric foreground galaxies for multiprobe analyses. The simplest method (conceptually) is a flux-limited selection, as applied in KiDS \citep[][hereafter \citetalias{Bilicki2018, Bilicki2021}]{Bilicki2018, Bilicki2021}. As shown in these papers, statistically accurate and precise photo-$z$ via an empirical (machine-learning) approach is possible for such a flux-limited foreground galaxy selection, provided that  the appropriate calibration (or, rather, training) for the data is available.

In this work, we present a method to further calibrate the overlap between spectroscopic training and photometric target data  and apply it to KiDS. In particular, we focus on the `KiDS-Bright' galaxy sample, which constitutes one of the foreground photometric datasets used in KiDS for GGL \citep{Brouwer2021, Georgiou2021, Burger2023KIDS, Dvornik2023KIDS}, the other being luminous red galaxies \citep{Vakili2019KIDS, Vakili2023KIDS}. Our approach can be used to assess the representativeness of spec-$z$ samples with respect to the photometric ones and also to improve the photo-$z$ precision of the latter; for instance by discarding those galaxies for which photo-$z$ performance is poor. We chose to perform such cleaning of galaxies in color-space, using a self-organizing map \citep[SOM,][]{kohonen1982self, Kohonenbook} by identifying galaxies in the photometric catalog that do not have counterparts in the available spec-$z$ sample.

Using KiDS Data Release 3 \citep[DR3,][]{deJong2017} and DR4 \citep{Kuijken2019}, respectively, \citetalias{Bilicki2018} and \citetalias{Bilicki2021} selected flux-limited galaxy samples at $r<20$ (mean $z\sim0.23$) with a negligible overall photo-$z$ bias and a scatter of $\sigma_{\delta z}\sim 0.018(1+z)$. The resulting KiDS-Bright dataset\footnote{Available for download at \url{https://kids.strw.leidenuniv.nl/DR4/brightsample.php}.} from the latter work covers the full DR4 footprint and contains roughly 1 million galaxies of surface density $\sim1000$ deg$^{-2}$. Both the selection of KiDS-Bright galaxies and the derived photo-$z$ were calibrated on the Galaxy And Mass Assembly spectroscopic survey \citep[GAMA,][]{Driver2011}. For its equatorial fields (which are fully covered by KiDS) GAMA offers close to 100\% complete spectroscopic measurements down to $r\sim19.8$ in SDSS Petrosian magnitude ($r_\mathrm{Petro}$). Thanks to the color-independent flux-limited galaxy selection both in GAMA and in KiDS-Bright, \citetalias{Bilicki2021} were able to obtain high-quality photo-$z$ in the latter not only for red galaxies but also for the blue ones, generally known to perform more poorly in this context. This was done by employing an artificial neural network approach \citep[ANNz2,][]{Sadeh2016ANN}, taking advantage of the very good match between the GAMA training set and the output KiDS-Bright sample. Furthermore, in a recent follow-up \cite{Anjitha2023} and John William et al (in prep.) improved these photo-$z$ further (reducing scatter by $\sim20\%$) by employing deep-learning methodologies.

At low redshifts, the lensing efficiency increases with the depth of the foreground sample. Therefore \citetalias{Bilicki2021}  chose to push to slightly fainter limits than allowed by GAMA completeness, taking the risk of including some galaxies not well matched to GAMA spectroscopy. Going slightly deeper than GAMA completeness was possible as that dataset includes a number of "filler'" targets at $r_\mathrm{Petro}>19.8$ \citep{Baldry2010}. Therefore,  \citetalias{Bilicki2021} could still train photo-$z$ rather robustly at the very faint end of KiDS-Bright. 

Even very accurate and precise photo-$z$ values may present variations with galaxy properties, such as color, magnitude, and type. Indeed, \citetalias{Bilicki2021} confirmed that red galaxies in KiDS-Bright have considerably better photo-$z$ than the blue ones. Some possible $r$-band magnitude dependence in photo-$z$ quality could also be observed at the faint end. Such dependencies often need to be identified and quantified for subsequent applications \citep[e.g.,][]{Burger2023KIDS}, while this may not always be possible if the spectroscopic calibration data do not fully represent the photometric sample in terms of magnitude or color.

In this work, we develop a method to quantify the completeness of the GAMA spectroscopic calibration data, for use in the photometric selection of KiDS-Bright foreground galaxies. We also show how a color-space comparison can be used to clean up the galaxy sample and remove objects that do not meet a given criterion, for instance, the photo-$z$ quality. For that purpose, we used the self-organizing map (SOM): a well-established tool for dimensionality reduction that allows us to project a multidimensional parameter space onto a two-dimensional (2D) manifold, while maintaining local associations between the objects from the original distribution in higher dimensional space. Unlike supervised machine learning, which uses spectroscopic data as the training dataset here in unsupervised machine learning, we used the photometric sample as the training dataset. Therefore, to avoid confusion, we refer to the spectroscopic data as the "calibration data." 

Our approach follows on  earlier applications in photometric surveys. \cite{Masters2015} utilized a SOM to identify photometric galaxies lacking spectroscopic representation for further observational follow-up, leading to the conception of the Complete Calibration of the Color-Redshift Relation (C3R2) Survey \citep{Masters2017C3R2, Masters2019}, carried out in the context of Euclid \citep{Euclid2021, Euclid2022}, 4-meter Multi-Object Spectrograph Telescope \citep[4MOST,][]{Gruen20234MOST}, and  Dark Energy Spectroscopic Instrument \citep[DESI,][]{mccullough2023desi}.
\cite{Wright2020SOM} developed a SOM-based framework to calibrate redshift distributions for weak lensing shear catalogs, and identify sources that were not represented in spectroscopic calibration data, resulting in a "gold" galaxy selection for weak lensing. This approach has been used extensively for KiDS analyses of cosmic shear  \citep[see, e.g.,][]{Wright2020KIDS450, Hildebrandt2021KIDS, Busch2022KIDS}. Furthermore, SOMs have also been utilized in the context of redshift calibration for DES \citep[see, e.g.,][]{Buchs2019,Myles2021}. 

In our study, we first show how SOMs can be used to compare color-space coverage between KiDS-Bright and GAMA. We then extended the spectroscopic sample by adding various surveys overlapping with the main KiDS footprint and several external `KiDZ' fields \citep{Wright2024}. As each SOM cell has associated average properties of a given sample (e.g., magnitudes, colors, redshifts), a comparison between, for instance, the mean photo-$z$ of KiDS-Bright and mean spec-$z$ of the calibration data can be used to eliminate those photometric galaxies where redshifts have been estimated poorly, similarly to what was done in the KiDS Gold selection \citep{Wright2020SOM}. Finally, by identifying those SOM cells where the spectroscopic representation in KiDS-Bright is still poor, we can select the photometric galaxies for possible future spectroscopic follow-ups, similar to the C3R2 survey.

The paper is organized as follows. Section \ref{s2:SOM} describes the terminologies of the self-organizing maps and their astrophysical application. In Sect.~\ref{s3:data}, we present the datasets utilized in this work. The projection of SOM and the completeness of GAMA with respect to KiDS is detailed in Sect.~\ref{s4:analysis}. This is followed by SOM comparison of the photometric and spectroscopic data  in Sect.~\ref{Sect:zcalib}  that allows us to remove the galaxies with worst-constrained photo-$z$s.  We present our conclusions and summary in Sect.~\ref{Sect:conc}. In the appendix, we discuss some further details of galaxy removal based on photo-$z$ performance. 

\section{Self organizing maps}
\label{s2:SOM}
A SOM is a form of artificial neural network utilizing unsupervised learning. The main goal of SOM is to transform high-dimensional input into a lower dimensional representation called a map or grid. This is achieved via the construction of an optimally representative manifold in $n$D, which can be subsequently visualized in two dimensions. The manifold can be planar or toroidal, the latter of which produces a map that wraps horizontally and vertically and is used in this study to avoid edge effects. Additionally, each cell within the map can be rectangular or hexagonal in shape: hexagonal cells being generally preferred, as they allow metrics (such as cell distance) to be computed between more neighbor cells (six compared to four for rectangular cells), thereby providing a finer and more detailed mapping of the input space. Thus, this study uses a toroidal map with hexagonal cells to enhance data representation and analysis.

\begin{figure}  
\centering
\includegraphics[height=7cm,width=8cm]{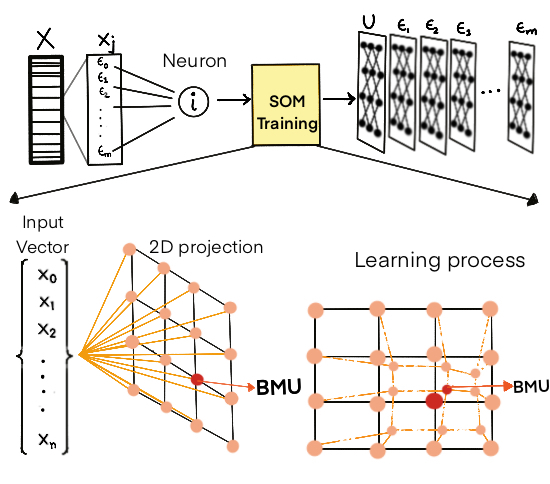}
\caption{Illustration of the projection of the $X$-input vector on a 2D SOM, based on Fig. 1 from \citet{SOM2020}. The red dot shows the best-matching unit.}
\label{Fig:fig_som}
\end{figure}

The SOM consists of nodes (or neurons) arranged in a 2D lattice, as shown in Fig.~\ref{Fig:fig_som}. Each node in the map is associated with a weight vector to represent the input patterns. At each training epoch, the algorithm computes the Euclidean distances between the input vector and the weight vectors of all neurons on the map. The node with the weight vector closest to the input vector (i.e., with the smallest Euclidean distance) is identified as the best-matching unit (BMU). The weight vector of this cell (and a subset of the map within the immediate vicinity of the BMU), are adjusted accordingly. During the training process, the learning rate controls the step size of weight adjustments and the cell shape defines the arrangement and connectivity of the nodes in the map, and the neighborhood function determines the extent of influence a BMU has on its neighboring nodes. The most commonly used neighborhood function is a Gaussian, which assigns weights to each node that are decreasing with its distance from the BMU according to the bell curve. As the training progresses, the gradual reduction of the radius of the neighborhood function focuses on smaller regions of the map, promoting finer adjustments and improving convergence. 

Figure~\ref{Fig:fig_som} demonstrates this process, whereby an input dataset $X$, containing $n+1$ sources, is used to train a SOM. Each source $X_{j}$ is tagged with $m+1$ parameters $\epsilon$, which are used to train the 2D map. After training, the sample can be visualized in one parameter (which may or may not have been used during training) by projecting said parameter onto the SOM.

SOMs have been shown to be particularly useful in calibrating redshift distributions and accessing color space match in the photometric and spectroscopic samples \citep{Carles2020, Wright2020KIDS450, Hildebrandt2021KIDS,  Oleksandra2021, Benjamin2023, Jafariyazani2023}. This is possible because, in addition to photometric data, features from the calibration spectroscopic sample can also be projected onto the SOM. Thanks to this, we can directly identify the cells occupied by both photometric and spectroscopic samples with similar properties. 

The cells showing significant deviations between photometric and spectroscopic redshifts indicate the types of galaxies for which the derived photo-$z$ are estimated poorly or the spec-$z$ is not well sampled in the cell. Identifying and removing such cells can help to refine the sample. They may also be used to single out potential candidates for spectroscopic follow-up to improve the parameter constraints. Therefore, the SOM can be utilized not only to identify a comprehensive representative calibration sample but also to discern clustering among similar galaxies within the feature space, based on their photometric data. 

In this work, we use the \texttt{SOMOCLU}\footnote{\url{https://somoclu.readthedocs.io/en/stable/download.html}} \citep[][]{somoclu2017} Python package for training SOMs on large data sets. We note that SOMs can be computationally intensive to train and SOMOCLU is a highly efficient, parallel, and distributed algorithm to train such maps. 

\section{Data}
\label{s3:data}
For our analysis, we combined photometric data originating from the VST and VISTA telescopes with spectroscopic redshifts taken from a number of surveys. Here, we provide details of the particular datasets. 

\subsection{Photometric data} 
\label{sect:photodata}
Our photometric data are based primarily on the Kilo-Degree Survey \citep[KiDS,][]{KiDS2013} observations and associated data products. KiDS\footnote{\url{http://KiDS.strw.leidenuniv.nl}} obtained optical imaging in four bands ($ugri$) with the OmegaCAM instrument mounted at the Cassegrain focus of ESO's VLT Survey Telescope \citep[VST,][]{Capaccioli2011} at Cerro Paranal, Chile. These optical data are combined with their infrared $ZYJHK_{\rm s}$ counterparts from the VIsta Kilo degree INfrared Galaxy survey \citep[VIKING,][]{Edge2013}, obtained using the Visible and InfraRed CAMera (VIRCAM) formerly mounted on ESO's 4m VISTA telescope also on Cerro Paranal. The combined KiDS+VIKING dataset is very well matched in terms of the depth and sky coverage \citep[e.g.,][]{Wright2019}.

Here we use KiDS Data Release 4 \citep[DR4,][]{Kuijken2019}, which covers roughly 1000 deg$^2$ and includes nine-band photometric information. The source detection was performed on $r$-band imaging using the \texttt{Source Extractor} \citep[][]{Bertin1996} software, which also provides Kron-like AUTO magnitudes. Images are also 
post-processed to measure the Gaussian aperture and PSF \citep[GAaP,][]{Kuijken2008} magnitudes. These were designed to yield accurate colors of galaxies \citep{Kuijken2015} and, hence, they are most appropriate for deriving color-sensitive quantities, such as photo-$z$ (see e.g., \citetalias{Bilicki2018}). Therefore, we used GAaP magnitudes also to build our feature space employed by the SOM. However, as discussed in  \cite{Kuijken2019}, for instance, GAaP magnitudes do not provide a good measurement of the total flux, especially for galaxies with large angular sizes, as in the case of KiDS-Bright. For the sample selection, we then followed \citetalias{Bilicki2021} and used the $r$-band  AUTO magnitudes to apply a flux limit. In the following, we have retained only the objects that have $r_\mathrm{auto} < 20$. We also require nine-band GAaP measurements to be available. To obtain a galaxy (i.e., extended source) selection, this is accompanied by the following cuts\footnote{See \cite{deJong2015} for details of these morphological flags.}: $\mathtt{CLASS\_STAR} < 0.5$, $\mathtt{SG2DPHOT} = 0$ and $\mathtt{SG\_FLAG} = 1$. The KiDS catalog also provides mask information, which encodes (in particular) the areas affected by various artifacts. We followed the general recommendations for the DR4 selection and removed objects with $(\mathtt{MASK} \& 28668) > 0$. Finally,  from the KiDS-Bright sample, we removed any objects that were assigned $z_\mathrm{phot}<0$ (a rare artifact of the ML model, usually highlighting issues with photometry). 

For our analysis, we supplemented the KiDS DR4 photometric data with optical+NIR measurements of similar quality as in KiDS+VIKING (KV) from the so-called "KiDZ" fields. These are areas mostly external to the KiDS footprint, which were targeted with VST and VISTA observations both by independent surveys and via dedicated observations of the KiDS team. These fields overlap with various deep spectroscopic surveys, which allows the KiDS team to calibrate redshift distributions with direct approaches (such as DIR, \citealt{Hildebrandt2017, Hildebrandt2020}, or SOM, \citealt{Wright2020KIDS450, Hildebrandt2021KIDS}) thanks to their having joint information of nine-band photometry and spectroscopy. For details of the KiDZ fields, see \citet{Wright2024}.  The data in the KiDZ fields include, for the vast majority of our sources, two separate measurements in the $i$-band to mimic the situation in KiDS DR5 \citep{Wright2024}. To homogenize these with the single $i$-band pass of DR4, we take the arithmetic mean of the magnitudes, which is a good approximation of mean flux for these bright objects with a high signal-to-noise ratio, especially given the two magnitudes are usually very consistent. If only one pass is available in KiDZ (a very rare occurrence, at a 0.02\% level), we used it as the $i$-band measurement. We also implemented masking and star removal, which  slightly differ from KiDS DR4  and are more in line with the DR5 post-processing. 

This study focuses on the KiDS-Bright sample (\citetalias{Bilicki2021}), a flux-limited galaxy dataset selected from KiDS DR4 as discussed above. The sample is accompanied by photo-$z$ generated with the neural network code ANNz2 \citep{Sadeh2016ANN}, where the training set was derived from a cross-match between the KiDS and GAMA equatorial datasets. These photo-$z$ are based on nine-band GAaP magnitudes and show very good statistical accuracy and precision, namely, mean bias (residual) $\langle \delta z \rangle= 5\times10^{-4} $ and scatter (scaled median absolute deviation from median, SMAD) $\sigma_{\delta z} = 0.018(1+z)$.  In the following, we compare these photo-$z$ with spectroscopic redshifts via a SOM projection. 

\begin{table*}
    \centering
    \caption{Surveys used in the extended spectroscopic compilation (Espec). } 
    \begin{tabular}{lrrrcl}
    \hline\hline
    Survey    & Median  & Median  & Count & Selection & References \\ 
              &   redshift     &   r-mag &       & criteria   &     \\
\hline
    GAMA-eq    &   0.220 &    19.16 &  145908 & $\mathtt{NQ}\geq3$, $z>0.002$ & \cite{Baldry2010} \\
    2dFGRS     &   0.117 &    17.85 &   48294 & $\mathtt{q\_z}\geq 3$, $z>0.002$ & \cite{Colless2001}\\
    GAMA-noneq &   0.207 &    18.96 &   41612 & $\mathtt{NQ}\geq3$, $z>0.002$ & \cite{Liske2015} \\
    SDSS DR14  &   0.152 &    17.54 &   30029 & $\mathtt{zWarning}=0$, $0<\mathtt{zErr}<0.001$, &  \\
               &         &          &         & $\mathtt{zErr}/z<0.01$, $z>0.001$ & \cite{Abolfathi2018} \\
    2dFLenS    &   0.229 &    18.64 &   19710 & $\mathtt{qual}<6$, $z>0.001$ & \cite{Blake2016} \\
    DESI EDR   &   0.261 &    19.82 &   12767 & \texttt{ZCAT\_PRIMARY}, $\mathtt{zWarn}=0$, $z>0.002$ & \cite{DESI2023} \\
    DEVILS     &   0.217 &    19.39 &     888 & $\mathtt{zBestType}=\mathtt{spec}$, $\mathtt{starFlag}=0$,  &  \\
               &         &          &         & $\mathtt{mask}=0$, $\mathtt{artefactFlag}=0$ & \cite{Davies2018} \\
    COSMOS &   0.221 &    19.49 &     675 & $3 \leq \mathtt{Q\_f} \leq 5$, or $13 \leq \mathtt{Q\_f} \leq 15$, or & \\
               &         &          &         & $23 \leq \mathtt{Q\_f} \leq 25$, or $\mathtt{Q\_f} \in \lbrace 6, 10 \rbrace$, $z>0.002$ & private comm. (M. Salvato)\\
    OzDES      &   0.200 &    18.72 &     516 & $\mathtt{qop} \in \lbrace3,4\rbrace$, $z>0.002$ & \cite{Lidman2020} \\
    WIGGLEZ    &   0.388 &    19.87 &     362 & $Q\geq3$, $z>0.001$, $Dz/z<0.1$ & \cite{Drinkwater2010} \\
    VVDS       &   0.228 &    19.55 &     212 & $\mathtt{ZFLAGS} \in \lbrace3,4,23,24\rbrace$ & \cite{LeFevre2005,LeFevre2013} \\
    HCOSMOS    &   0.222 &    19.50 &     178 & none & \cite{Damjanov2018} \\
    ACES       &   0.181 &    19.29 &     173 & $\mathtt{Z\_QUALITY} \geq 3$, $\mathtt{zErr}/z<0.01$ & \cite{Cooper2012} \\
    G15-DEEP   &   0.250 &    19.91 &     106 & $\mathtt{Z\_QUAL}\geq 3$, $z>0.001$ & \cite{Driver2022GAMA} \\
    \hline\hline
\end{tabular}
\begin{tablenotes}
\item The table is arranged based on the descending number of sources within the KiDS DR4 + KiDZ bright sample coverage area. All the numbers apply after cross-matching the input spec-$z$ samples with the photometric data limited at $r<20$ mag and removing duplicates between surveys. We list only those surveys that have at least 100 cross-matched objects each.
\end{tablenotes}
\label{tab:survey}
\end{table*} 

In addition to the photo-$z$, in \citetalias{Bilicki2021} a number of physical galaxy properties were derived for the KiDS-Bright dataset by employing the LePhare code \citep{Arnouts1999, Ilbert2006} on the KV fluxes. Here, we use one of these properties (galaxy stellar masses), which were shown in \citetalias{Bilicki2021} to be consistent with the more accurate derivations from GAMA spectroscopy \citep[e.g.][]{Taylor2011}. Stellar masses are relevant for our study, as these were employed by \cite{Dvornik2023KIDS} to bin the lens sample extracted from KiDS-Bright and to link the observed clustering and lensing signal of galaxies with the theoretical framework of the halo model (via the conditional stellar mass function).    

\subsection{Spectroscopic data}
\label{Sect:specdata}
Our methodology relies on combining photometric data with redshift measurements. We obtain the latter from a number of spectroscopic surveys and datasets overlapping with KiDS DR4 and the KiDZ fields. 

\subsubsection{GAMA}
\label{sect:gama}
The basic spectroscopic dataset we use is derived from the Galaxy And Mass Assembly \citep[GAMA,][]{Driver2009} final Data Release 4 \citep{Driver2022GAMA}. The GAMA survey observed $\sim$ 300,000 galaxies down to $r \lesssim 19.8$ over 286 deg$^2$ using the AAOmega multi-object spectrograph on the 4m Anglo-Australian Telescope. This includes five fields, of which four are fully contained within KiDS: three equatorial fields \citep[G09, G12 and G15;][]{Baldry2010}, and one southern field (G23). The last GAMA field \citep[G02,][]{Liske2015} partly overlaps with KiDZ. In addition, a dedicated small-area, deeper survey lies within the G15 field \citep[G15-Deep,][]{Driver2022GAMA}. Of these, the most spectroscopically complete are the equatorial fields (GAMA-eq hereafter), where redshift targets were originally selected from SDSS as flux-limited to $r_\mathrm{Petro}<19.8$ (with no color preselection) plus some additional filler targets fainter than this limit. The estimates of GAMA-eq completeness were originally at the level of $98.5\%$ for its fiducial flux limit \citep{Liske2015}. However, subsequent adoption of KiDS photometry for GAMA galaxies \citep{Bellstedt2020GAMA} led to this being revised to $98\%$  at $r_\mathrm{KiDS}<19.6$ \citep{Driver2022GAMA}, where `KiDS' refers to flux measurements made with the code ProFound \citep{Robotham2018}. \citet{Bellstedt2020GAMA} then revised this completeness estimate to 95\% at $r_\mathrm{KiDS}<19.72$  \citep{Driver2022GAMA}. 

The GAMA-eq sample was used to calibrate the KiDS-Bright selection and photo-$z$ estimation in \citetalias{Bilicki2021}, and we will employ it as the main spectroscopic reference for our study. The other GAMA fields (G02 and G23; GAMA-noneq hereafter), have brighter flux limits and much less complete sampling than GAMA-eq \citep{Liske2015}, and only serve as extensions together with other spec-$z$ samples discussed below. In all cases, GAMA redshifts were selected with the quality flag $\mathrm{NQ} \geq 3$ and with $z>0.002$ to avoid stellar contamination.
\subsubsection{Extended spectroscopic compilation}
\label{sect:especz}
In addition to the GAMA data detailed above, we used a number of other spectroscopic datasets overlapping with KiDS DR4 and KiDZ. Most of these are the same as used in KiDS cosmic shear studies and detailed in previous papers \citep{Wright2024,Busch2022KIDS}. However, we also added a few others. Also, most importantly, our cuts on the photometric data are different. The properties of these additional redshift samples are provided in Table \ref{tab:survey}. The numbers and statistics in the table are applicable after a cross-matching with KiDS/KiDZ photometric data, cutting down to $r_\mathrm{auto}<20$ to mimic the KiDS-Bright selection and keeping only unique objects\footnote{The matching is done within $1^{\prime\prime}$ radius. See \citet{Wright2024} for details on how duplicates are handled.}. The datasets are listed in descending order of the number of sources within KiDS DR4 + KiDZ after the KiDS-Bright selections have been applied. We  denote this extended and matched spectroscopic compilation as `Espec'.

As detailed in Table~\ref{tab:survey}, the largest contributions in terms of total numbers of galaxies come from wide-angle surveys overlapping with KiDS or KiDZ. After requiring our $r<20$ selection, most of these surveys become effectively shallower than GAMA, even if they originally included deeper, higher-redshift sources. This is because, as in the case of SDSS or 2dFLenS, the higher-$z$ sources are typically color-preselected and/or sparsely sampled, while the more complete, magnitude-limited samples typically employ magnitude limits brighter than used by GAMA. The remaining contributions to our extended compilation are from small-area, deep surveys, which typically have (considerably) larger mean redshifts than GAMA, even after our flux limit is applied. However, these samples have very limited overlap with our photometric data, given their small on-sky areas. Finally, in between these extremes, there is the DESI Early Data Release (EDR), which is generally deeper than the wide-angle surveys and spans a relatively large area on-sky. However, the DESI EDR only intersects with approximately 44 sqdeg of the KiDS/KiDZ footprint, which limits the influence of these spectra. Future releases from DESI, with greater intersection with KiDS, will greatly improve the influence of DESI in analyses of KiDS data such as ours.

\section{Color space analysis of KiDS-Bright galaxies}
\label{sect:som_app}
In this section, we use the SOM to compare the consistency between the color-spaces of the KiDS-Bright sample and the available spectroscopic sources, to evaluate how complete and/or representative the latter are with respect to the former. Initially, we examine the completeness of the GAMA-eq data in comparison to KiDS-Bright. Subsequently, we explore the potential of extending the spectroscopic sample to aid in selecting subsamples with the most reliable photo-$z$.

We trained a $30 \times 30$ hexagonal-cell SOM with toroidal topology using the KiDS-Bright photometric sample of $\sim$1 million galaxies. This gave us on average $\sim1000$  galaxies per cell, which provides sufficient sampling per cell for our analysis. We trained our SOM using nine magnitudes and 36 colors as our feature space and run the training for 100 epochs, which gave an optimal balance between acceptable map convergence and maintaining reasonable computational runtime. The learning scale was initialized with a value of 0.1 and is gradually reduced to 0.01 in the final epoch. The SOM, therefore, provides us with a 2D projection of the full 45 dimensions of KiDS-Bright color-magnitude space, which we utilized for our subsequent analyses. Furthermore, we also mapped various other parameters not used in the training of our SOM, including details of the spectroscopic samples discussed previously in Sect.~\ref{Sect:specdata}.

\begin{figure*}
\centering
\includegraphics[width=0.245\textwidth]{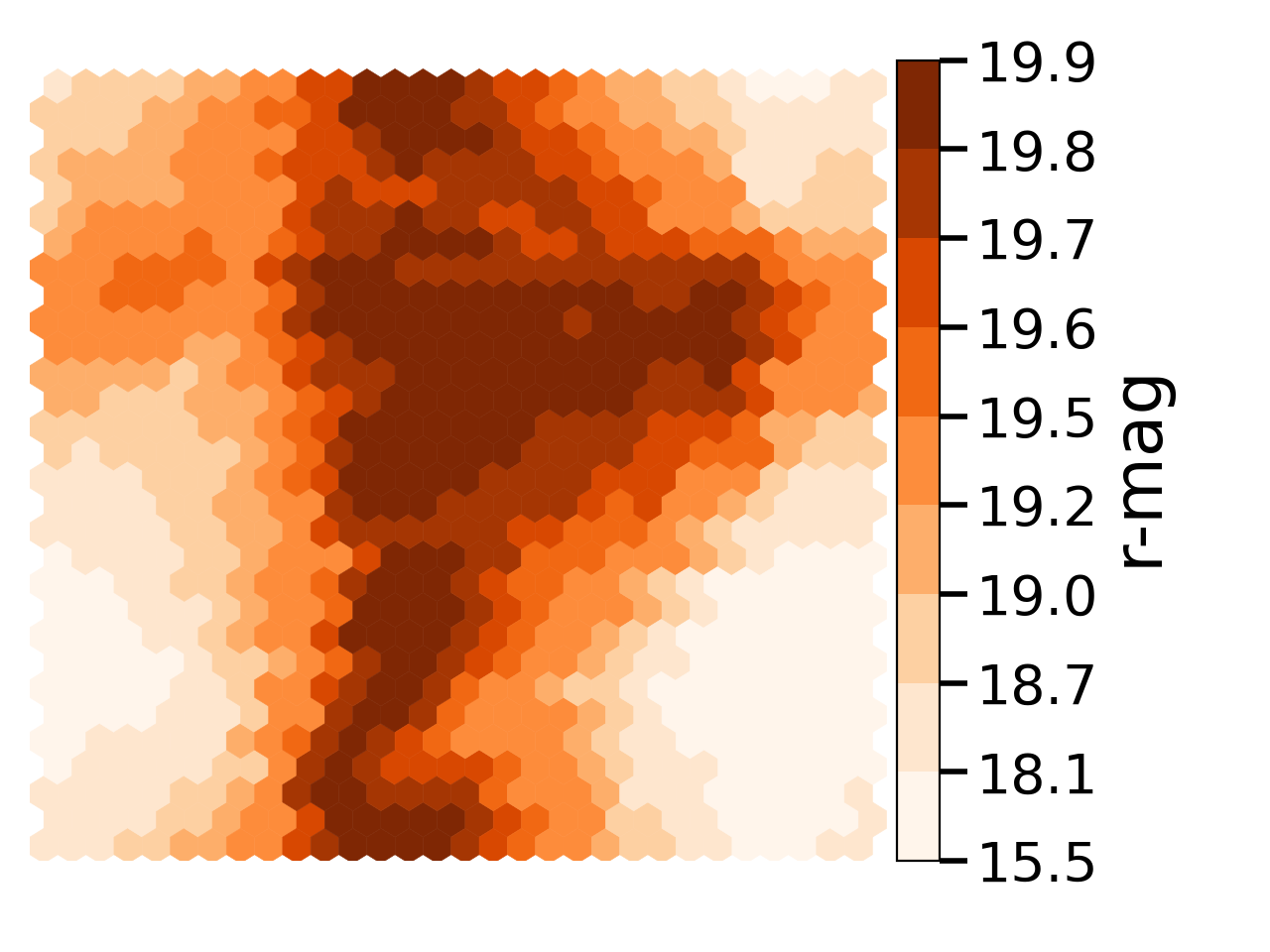}
\includegraphics[width=0.245\textwidth]{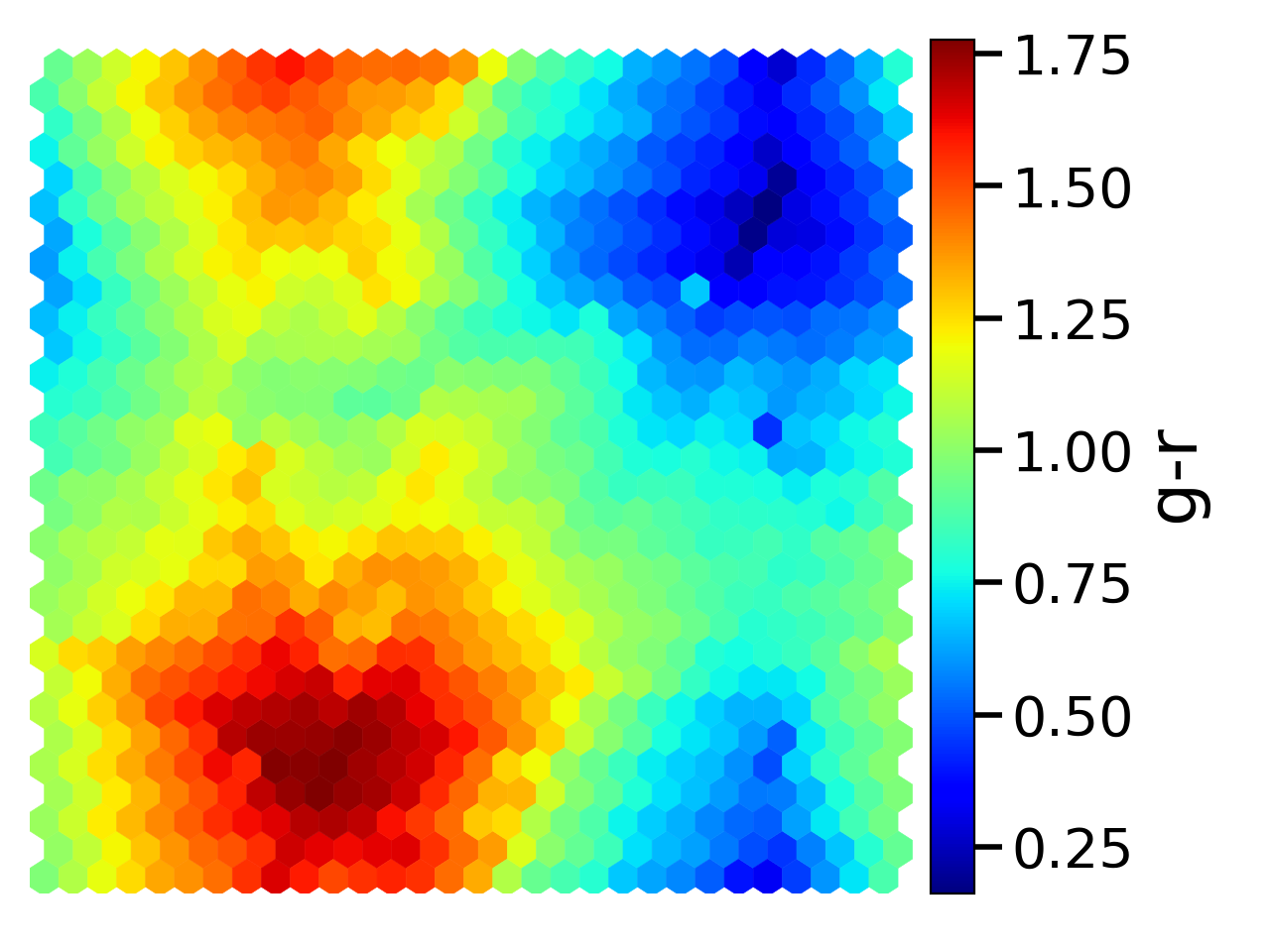}
\includegraphics[width=0.245\textwidth]{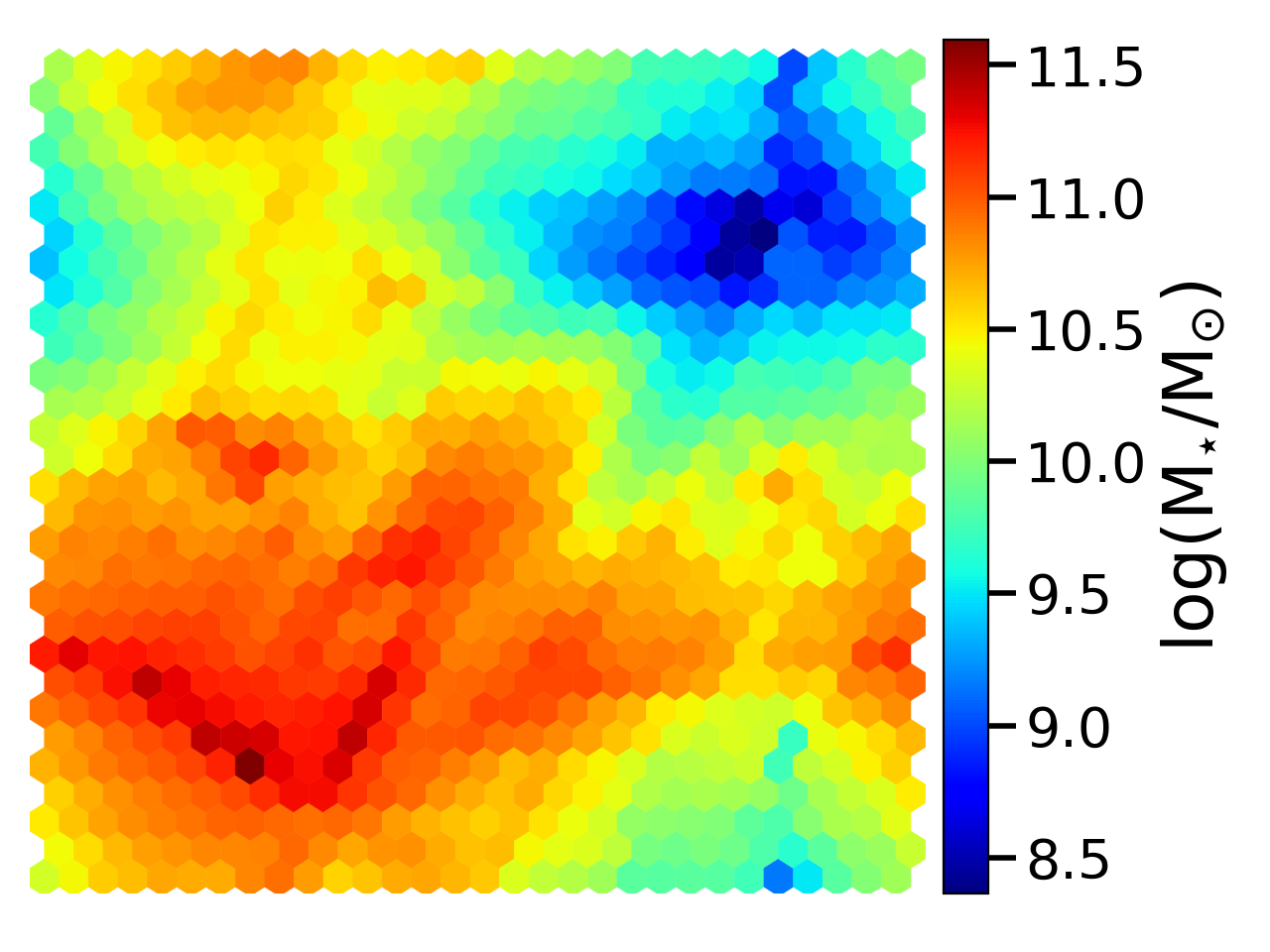}
\includegraphics[width=0.245\textwidth]{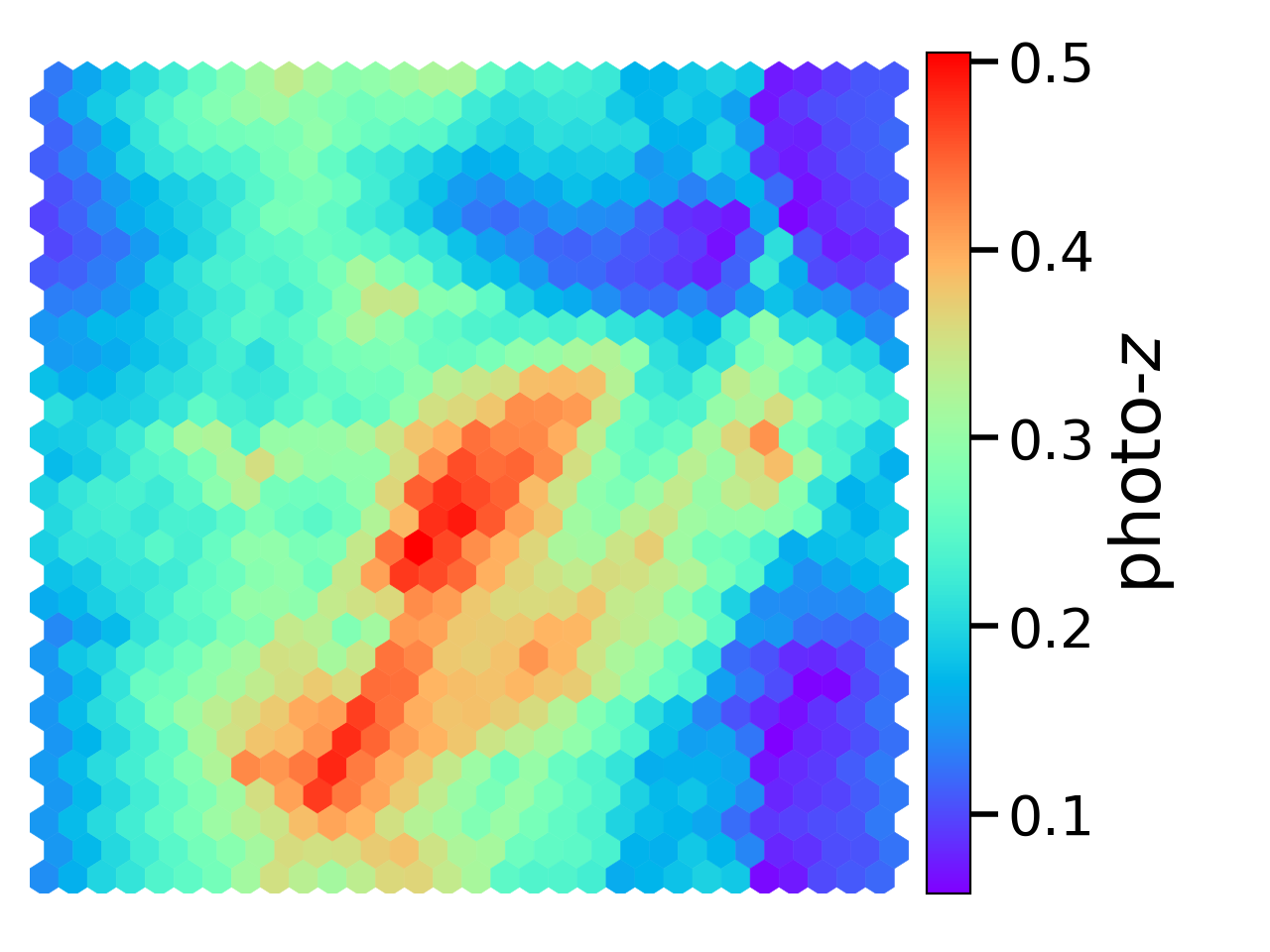}
\caption{ 30 $\times$ 30 hexagonal-cell SOMs with toroidal
topology trained using the KiDS-Bright photometric sample (9 magnitudes+36 colors) of
$\sim$1 million galaxies. These plots show a few example properties of the KiDS-Bright photometric sample projected on the SOM. From left to right: (1) $r$-band AUTO magnitude, (2) $g-r$ color, (3) logarithm of stellar mass in solar units, and (4) photometric redshift derived with ANNz2. Properties (1) and (2) are taken from the KiDS DR4 photometric dataset \citep{Kuijken2019}, while (3) and (4) were obtained for the KiDS-Bright sample by \citetalias{Bilicki2021}. Colors indicate the mean value of a given property per SOM cell.}
\label{Fig:rKiDS}
\end{figure*}

\begin{figure*}
\centering
\includegraphics[width=0.32\textwidth]{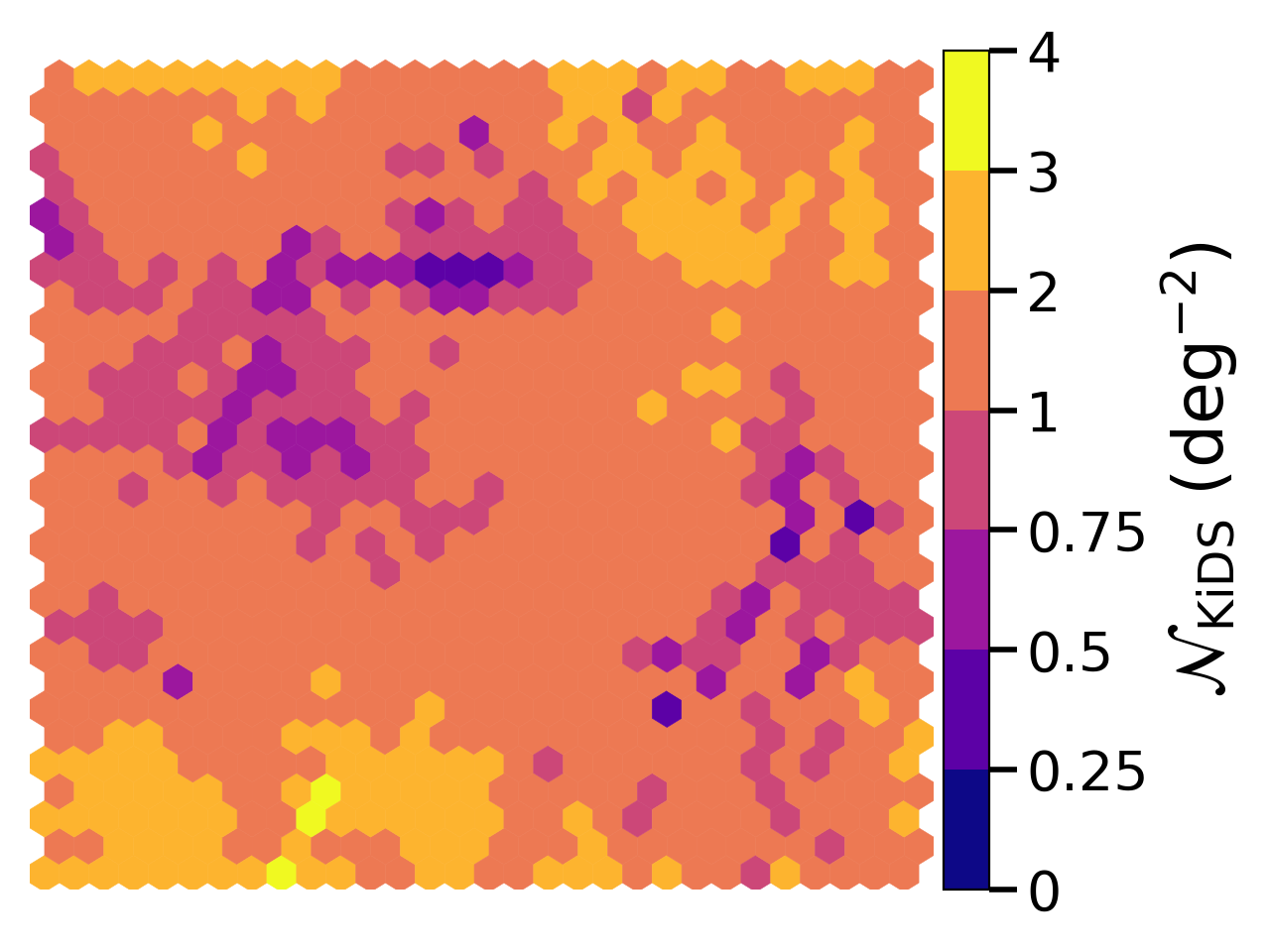}
\includegraphics[width=0.32\textwidth]{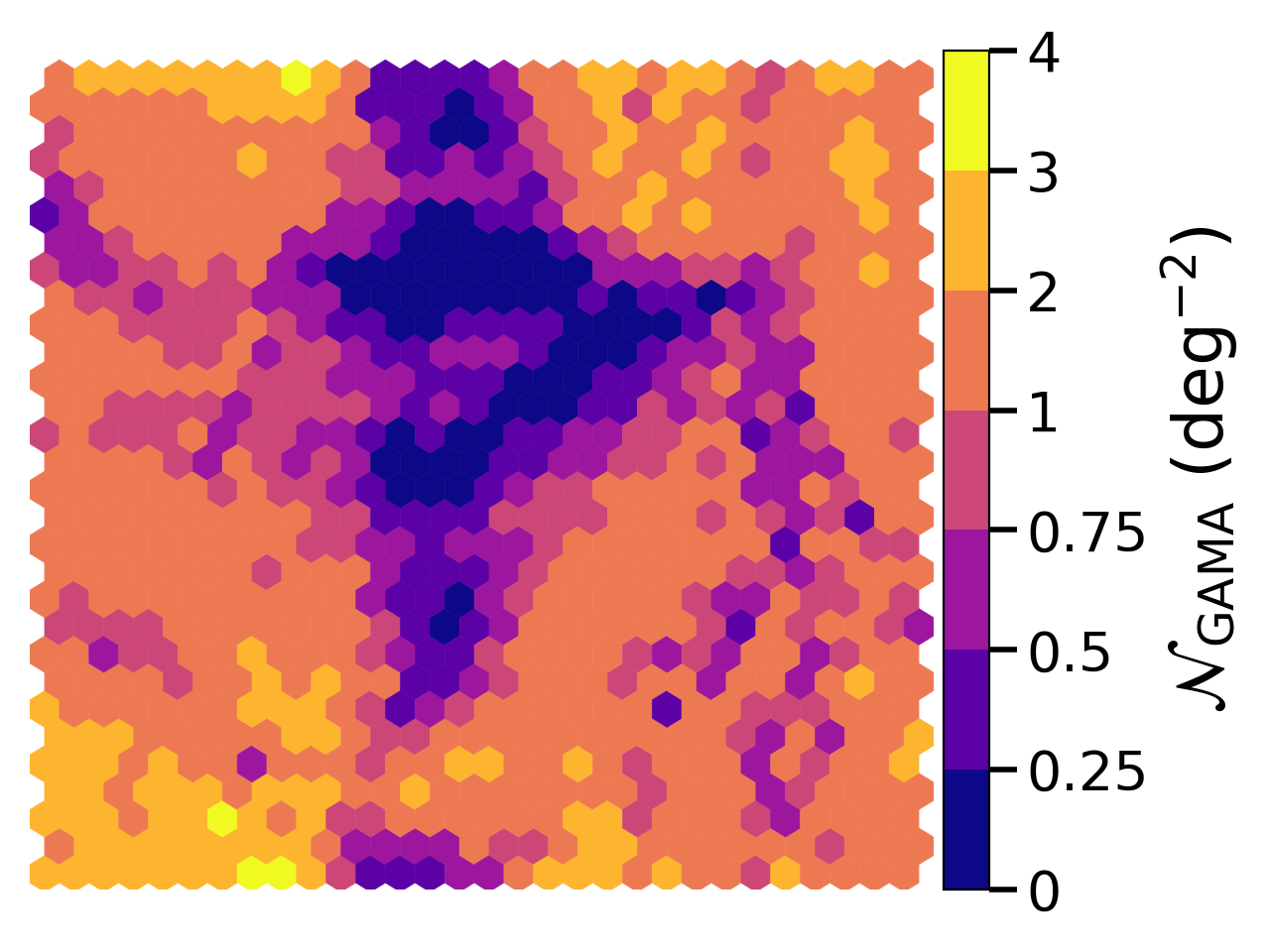}
\includegraphics[width=0.32\textwidth]{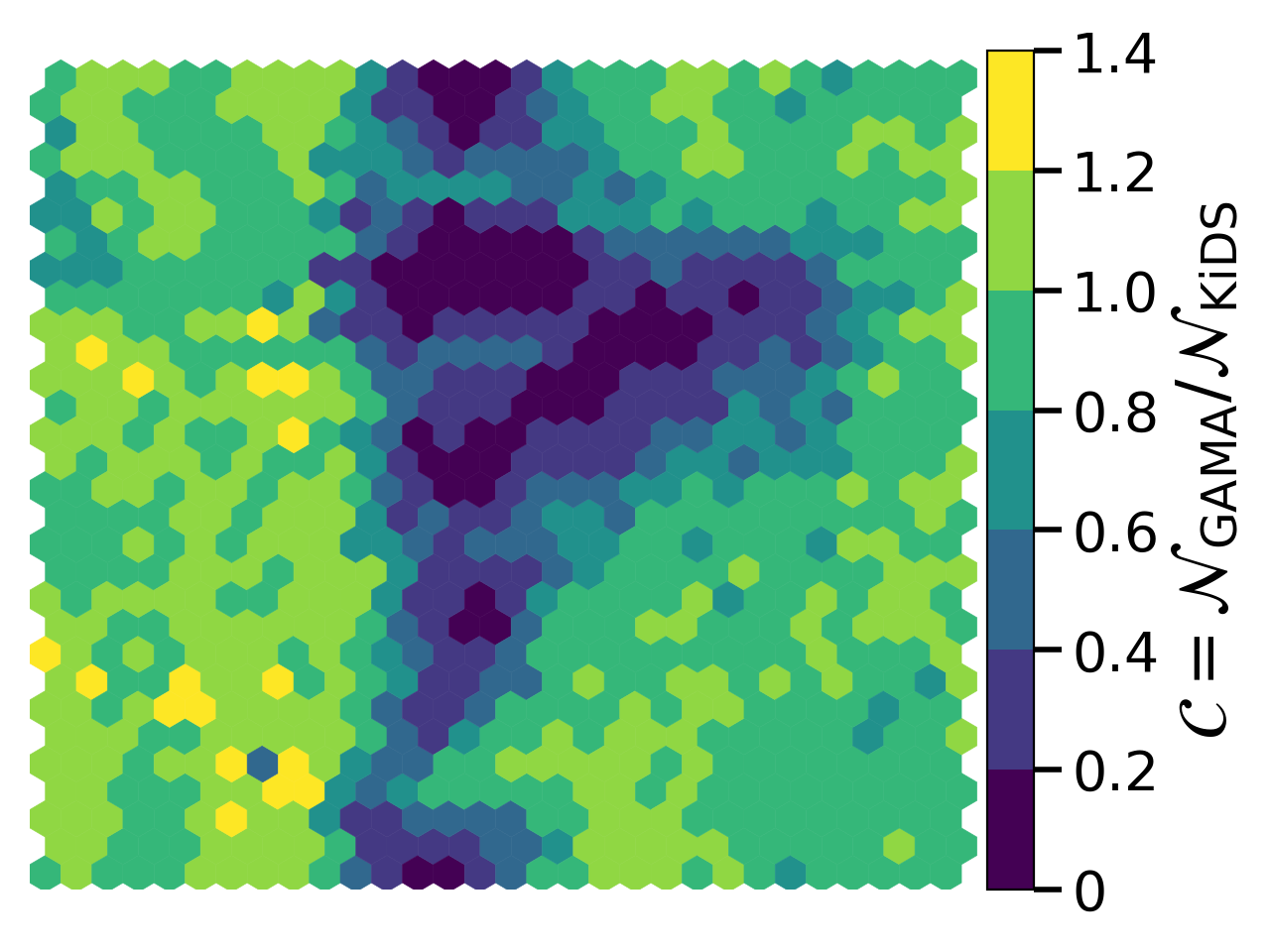}
\caption{Comparison of the number density of the KiDS-Bright photometric sample (left), GAMA spectroscopic (middle), both expressed as average sky-projected density per SOM cell and the ratio of the latter to the former (right), giving a measure of GAMA completeness with respect to KiDS-Bright. The ratio above unity can appear as a result of the cosmic variance between GAMA and KiDS datasets.}
\label{Fig:count}
\end{figure*}

\begin{figure*}
\centering
\includegraphics[width=1\textwidth]{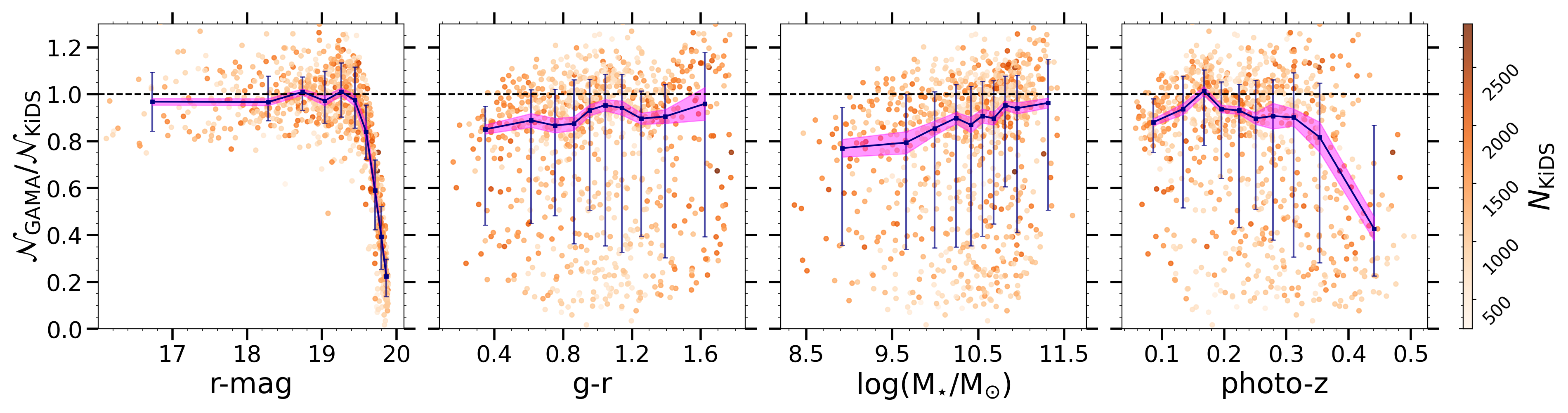}
\caption {Dependence of the GAMA completeness (w.r.t. KiDS) on selected properties of the KiDS-Bright sample, based on SOM projections. The completeness is expressed as the per-cell ratio of GAMA and KiDS-Bright surface density. The panels show from left to right: $r$-band AUTO magnitude, $g-r$ color, logarithm of stellar mass, and  ANNz2 photometric redshifts. Each orange dot represents one SOM cell, while the blue points are median values with central 68-percentile error bars. Binning is done by preserving the same number of averaged cells per bin. The pink-shaded regions show the bootstrap error from 10000 bootstrap samples, as an approximation to the uncertainty on the median.}
\label{Fig:ratio_rzug}
\end{figure*}

\subsection{Comparison of KiDS-Bright and GAMA}
\label{s4:analysis}
We start by analyzing how various KiDS-Bright galaxy properties are mapped on the SOM. This allows us to make a qualitative comparison of how these quantities are correlated, including both the features employed to train the SOM and additional ones. In Fig.~\ref{Fig:rKiDS}, we show four of them: $r$-band magnitude (first panel), $g-r$ color (second panel),  stellar mass (third panel), and photometric redshift (fourth panel). The two former are taken directly from the observations \citep{Kuijken2019}, the two latter were derived from KiDS+VIKING photometry as described in \citetalias{Bilicki2021}. Stellar masses ($M_*$) were obtained using LePhare \citep{Arnouts1999, Ilbert2006}, and photo-$z$ were estimated using ANNz2 \citep{Sadeh2016ANN}. Finally, we hereafter refer to $r$-band AUTO magnitudes simply as $r$ magnitudes, while colors are based on GAaP photometry. 

The visualization reveals distinct feature clustering patterns. Fainter galaxies tend to gather in the central region of the map, while brighter ones cluster towards the edges (which are interconnected in toroidal mapping). Additionally, as expected, we see segregation in color space -- one group tending towards bluer and the other towards redder. We see similar trends in all the nine optical and infrared bands and the corresponding 36 colors. Remembering that the respective cells in different panels of Fig.~\ref{Fig:rKiDS} correspond to the same sets of galaxies, we can observe a correlation between galaxy color and stellar mass. Moreover, at higher redshifts, galaxies of predominantly higher stellar masses, as well as redder and fainter, are prevalent, whereas at lower redshifts, a wider distribution of various stellar masses and colors is evident. All these patterns are in line with the expected properties of a flux-limited sample, such as KiDS-Bright.

Next, we make a direct comparison between KiDS-Bright and GAMA using our SOM. To reiterate, the GAMA equatorial spectroscopic sample was utilized by \citetalias{Bilicki2021} both to calibrate the selection of KiDS-Bright and to derive its photo-$z$ through an ML approach. Consequently, our objective here is to examine whether, and to what extent, GAMA-eq might exhibit incompleteness (and/or non-representativeness) in relation to the KiDS-Bright galaxies. In the subsequent paragraph, we will discuss how this potential incompleteness could impact the accuracy of KiDS-Bright photo-$z$.

We start by noting that the projections from Fig.~\ref{Fig:rKiDS} look very similar between KiDS-Bright and GAMA, so we are not showing them here for the latter sample. However, more striking differences start to appear if we move to other quantities. An example is the number density ($\mathcal{N}$), which we obtain by dividing the galaxy counts of each SOM cell by the effective survey area (777 deg$^2$ for KiDS-Bright and 142 deg$^2$ for GAMA). The result is plotted in Fig.~\ref{Fig:count} (left and middle panels). By comparing these with the left-hand side panel of Fig.~\ref{Fig:rKiDS}, we see that the number density in the GAMA survey is much lower than in KiDS-Bright at the faint end of the latter. This is expected because the adopted KiDS-Bright flux limit of $r_\mathrm{auto}<20$ mag was slightly fainter than the GAMA-eq completeness limit of $r_\mathrm{petro}\lesssim19.8$. 

The right-hand panel of Fig.~\ref{Fig:count} illustrates the per-cell completeness defined as $\mathcal{C} \equiv \mathcal{N}_\mathrm{GAMA}$/$\mathcal{N}_\mathrm{KiDS}$. Here, completeness implies the parametric completeness of GAMA with respect to KiDS in magnitude. We note that this can be occasionally above unity, as for a given SOM training, the number of GAMA galaxies in some cells can be higher than that number for KiDS. This is due to two reasons: GAMA has more representative galaxies in a given cell, and also due to cosmic variance between GAMA and KiDS datasets. The faint-end incompleteness of GAMA, visible in the center of the middle and right panels of Fig.~\ref{Fig:count}, does not need to mean that the KiDS-Bright photo-$z$ derived in \citetalias{Bilicki2021} are unreliable there, provided that the GAMA selection at $r\sim20$ mag seems to sparsely sample the KiDS-Bright color space  (Fig.~\ref{Fig:rKiDS}). 
\begin{figure}
\centering
\includegraphics[width=0.48\textwidth]{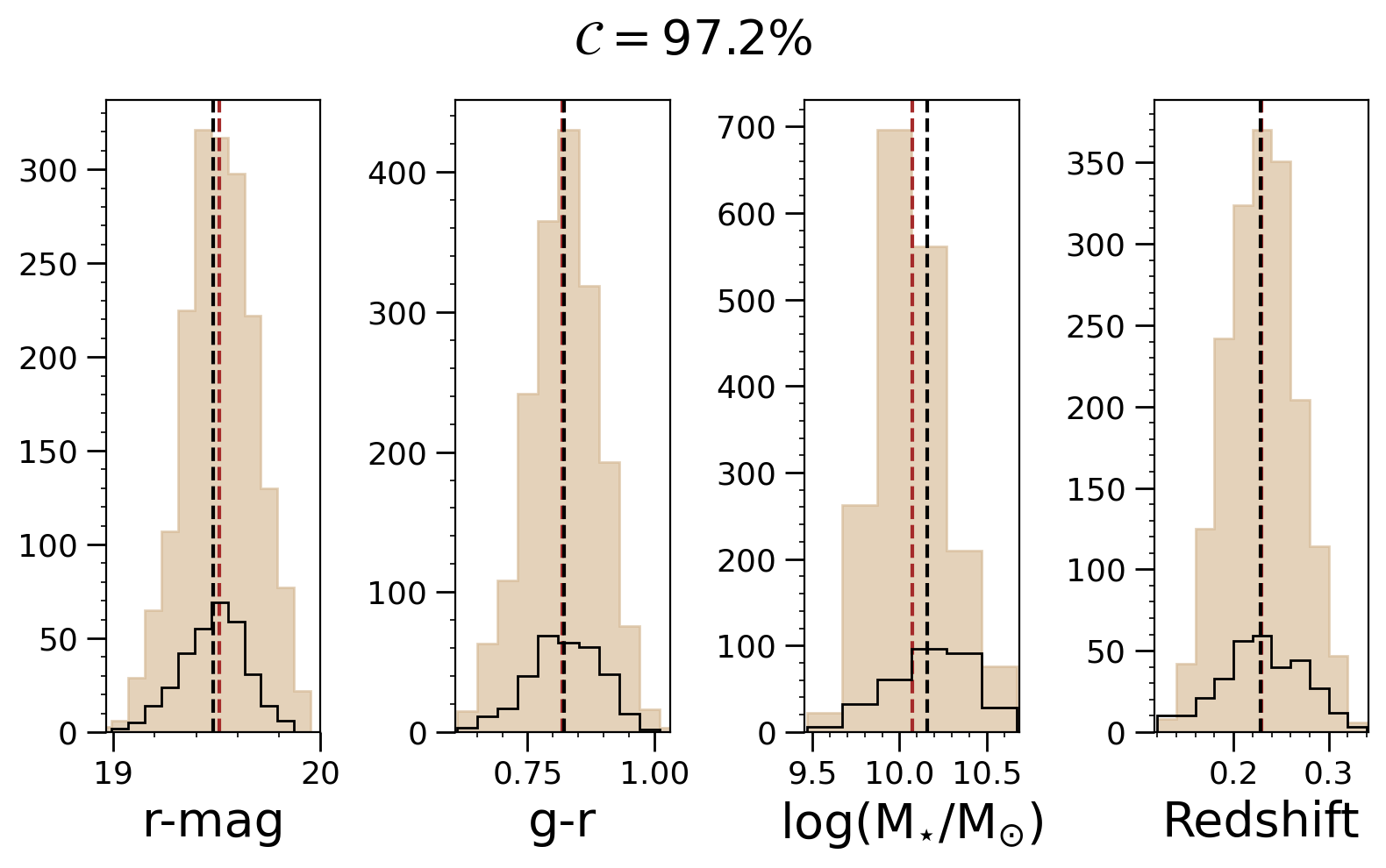}
\includegraphics[width=0.48\textwidth]{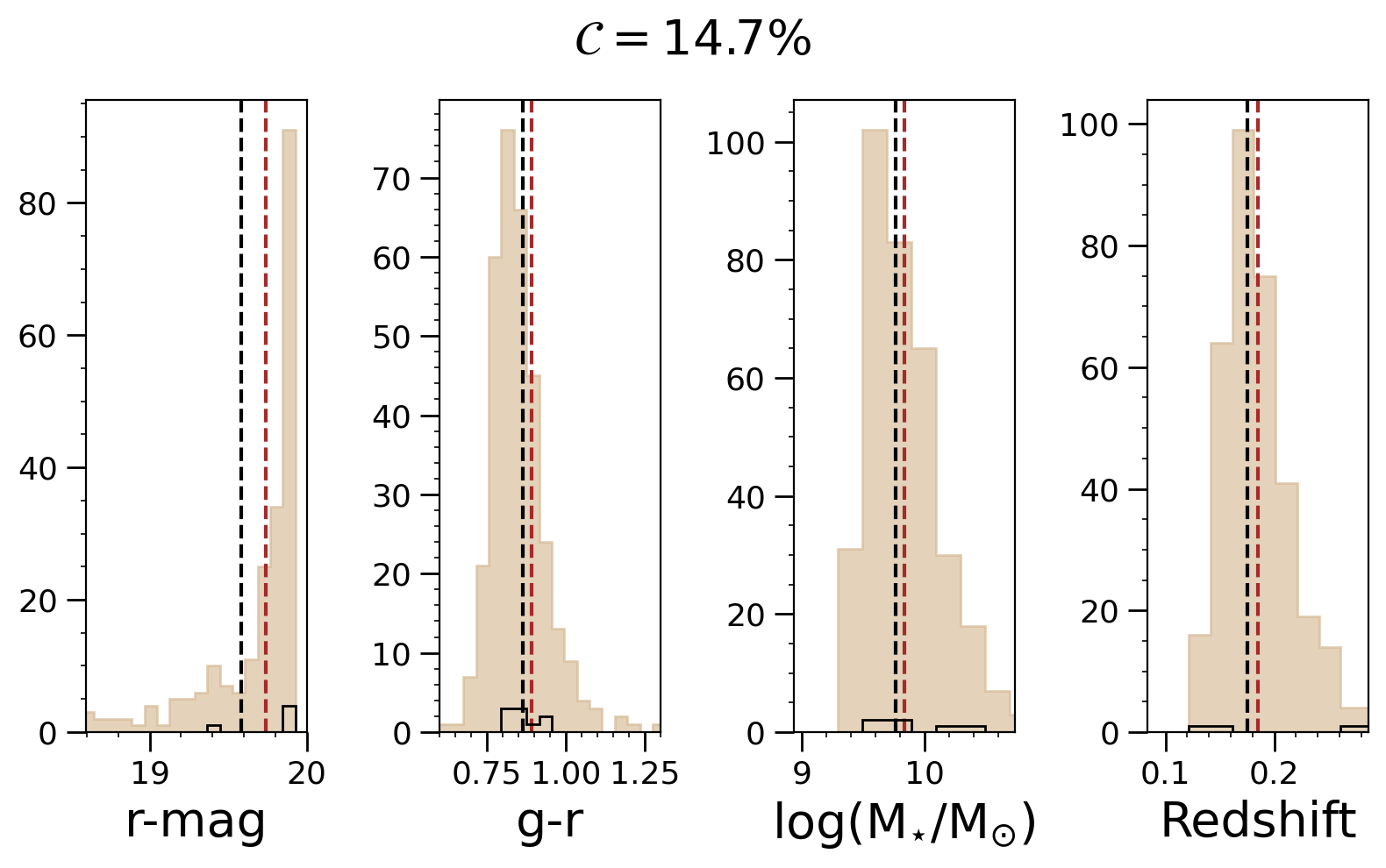}
\caption{Distributions of selected galaxy properties for two example SOM cells, one where GAMA completeness is very high (cell [18,29], upper panels) and one where it is low (cell [15,22], lower panels). From left to right, we show the observed $r$-band magnitude, $g-r$ color, stellar mass, and ANNz2 photometric redshift. Photometric KiDS-Bright data is plotted as filled brown bars, while spectroscopic GAMA is given by black lines. Vertical dashed lines indicate means: for KiDS-Bright in red and for GAMA in black.}
\label{Fig:cel37_prop}
\end{figure}

The completeness, $\mathcal{C}$, measured from the right-hand panel of Fig.~\ref{Fig:count}, can be directly mapped to other properties through the use of the SOM. Examples are provided in Fig.~\ref{Fig:ratio_rzug}, where we show the completeness as a function of the same quantities as visualized in Fig.~\ref{Fig:rKiDS}. Each individual point represents one SOM cell, and we see considerable scatter among these. This scatter is the result of a combination of sample variance, shot noise, and selection effects in the two samples. However, by averaging over the individual cells in bins of the respective $x$-axes, we can trace the overall patterns in the (in)completeness of GAMA-eq vs. KiDS-Bright. We illustrate this with the blue points indicating medians with the error bars reflecting the $(16,84)^{\rm th}$ percentile range.

On the one hand, GAMA keeps its 98\% completeness up to $r\sim 19.5$. Taking into account the differences between the AUTO magnitudes we use and those from \texttt{ProFound} employed by GAMA \citep{Bellstedt2020GAMA}, our findings are consistent with the 19.6 mag value in \cite{Driver2022GAMA} for the same completeness level. Beyond that range, the ratio plummets, in line with the qualitative assessment we made above. Looking at the same as a function of color, no clear trend emerges, though, except for perhaps a suggestion of GAMA being a bit less complete for blue rather than for red galaxies, driven by the fact that faint things are preferentially blue. Similar trends with respective apparent magnitudes and colors are seen for other bands (not shown here). More pronounced trends in GAMA incompleteness emerge in KiDS-Bright stellar masses and photo-$z$, although it is worth noting that, except for the most extreme bins (respectively $M_*\sim 10^9 M_\odot$ and $z_\mathrm{phot}\sim0.4$), the $\mathcal{N}_\mathrm{GAMA}$ to $\mathcal{N}_\mathrm{KiDS}$ ratios are consistent with unity within 68-percentiles. An interesting observation is that individual incomplete cells, which are practically all at the faint end (see the leftmost panel in Fig.~\ref{Fig:ratio_rzug}), are scattered all over the panels if we consider other quantities than magnitudes. Still, it is reassuring to see that the completeness stays (well) above 80\% independently of, e.g., the $g-r$ color. This could be one of the reasons why in \citetalias{Bilicki2021}, the photo-$z$ performance was robust for both red and blue galaxies, even beyond the GAMA completeness limit.

Finally in Fig.~\ref{Fig:cel37_prop} we highlight two SOM cells: one boasting over 97\% GAMA completeness and the other at the lower end, of approximately 15\%. We depict per-cell histograms of the same quantities as presented earlier in Figs.~\ref{Fig:rKiDS} and~\ref{Fig:ratio_rzug}. The filled brown bars represent KiDS-Bright, while the solid black lines denote GAMA; vertical dashed lines indicate the mean. The upper panel illustrates an instance of very high completeness, where the mean values of the respective quantities overlap between KiDS-Bright and GAMA, which should result in a very good correspondence between the estimated photo-$z$ and GAMA spec-$z$. However, even in cases of low completeness, depicted in the bottom row, the photo-$z$ estimates also align well with the GAMA spec-$z$ in the cell. Magnitudes and colors are utilized both in the SOM training and in deriving photometric redshifts, and hence it is expected that these training parameters would be consistent across each cell, resulting in similar derived parameters. These examples underscore the earlier findings of unbiased photometric redshift measurements in \citetalias{Bilicki2021} owing to GAMA completeness and representative sampling in color.

\subsection{Selection of a clean sample using SOM}
\label{Sect:zcalib}
In the previous section, we quantified the relative completeness of the GAMA spectroscopic data with respect to the photometric KiDS-Bright as a function of several galaxy properties. We observed that the GAMA completeness quickly drops at the faintest and highest-$z$ end, while no significant deterioration is present with regard to colors or stellar mass. As GAMA was used to train the KiDS-Bright photo-$z$, here we discuss how the SOM comparison of the photometric and spectroscopic data could be used to improve GAMA representation with respect to KiDS-Bright. As a result, we may remove galaxies with ill-constrained photo-$z$ from the KiDS-Bright sample, as determined by direct comparisons with both GAMA itself and the extended spectroscopic sample (Espec) presented in Sect.~\ref{sect:especz}.

As the left-most panel of Fig.~\ref{Fig:ratio_rzug} suggests, the simplest way to improve GAMA representation would be to lower the magnitude limit of KiDS-Bright (raise the flux threshold), as the majority of the incomplete SOM cells are at the faintest end. This would, however, affect the usefulness of the photometric sample for lensing studies by decreasing its effective depth, while it would not necessarily lead to direct improvement in photo-$z$ quality, as redshift estimates depend on the full nine-band information. We therefore compare such a more aggressive flux cut with SOM-based analysis of photo-$z$ quality, where we will consider both relative and absolute photo-$z$ errors as our test statistics. The former is defined as
\begin{equation}
    Dz\equiv \delta z / z_\mathrm{spec}
    \label{eq:rel}
,\end{equation} namely, as a fractional difference,  expressed in percent, where $\delta z \equiv z_\mathrm{phot} - z_\mathrm{spec}$. As photo-$z$ always have non-negligible uncertainties, including at very low redshift, this relative error $Dz$ is expected to diverge as $z \rightarrow 0$, but otherwise be stable for most of the redshift range. Our second statistic, the absolute photo-$z$ error, is defined as 
\begin{equation}
\Delta z \equiv \delta z / (1+z_\mathrm{zspec}),
  \label{eq:abs}
\end{equation} 
where we account for the fact that photo-$z$ errors usually scale with $1+z$. This metric should be relatively constant with photometric redshift, but will vary with true spectroscopic one as well as galaxy type.

\begin{figure}
\centering
\includegraphics[width=0.22\textwidth]{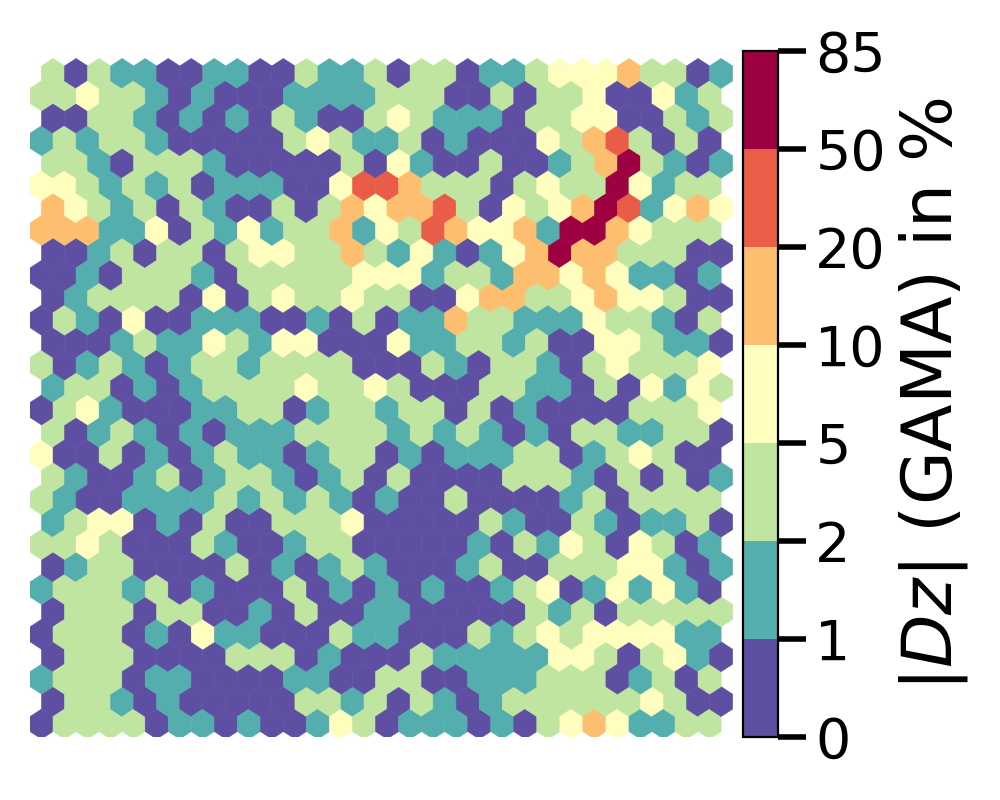}
\includegraphics[width=0.22\textwidth]{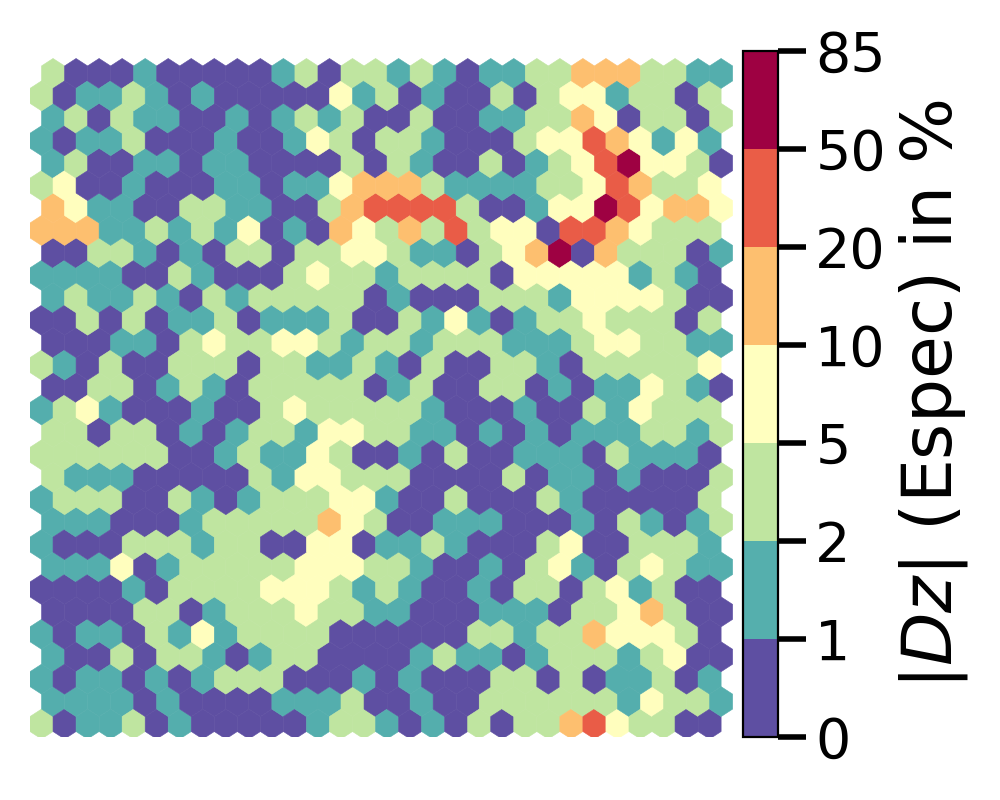}
\caption {The relative difference $Dz \equiv \delta z / z_\mathrm{spec}$ between the KiDS-Bright photometric and calibration spectroscopic redshifts, calculated in each SOM cell. The left and right panel shows  $|Dz|$ for GAMA and Espec, respectively. }
\label{Fig:zrel}
\end{figure}

\begin{figure}  
\centering
\includegraphics[width=0.22\textwidth]{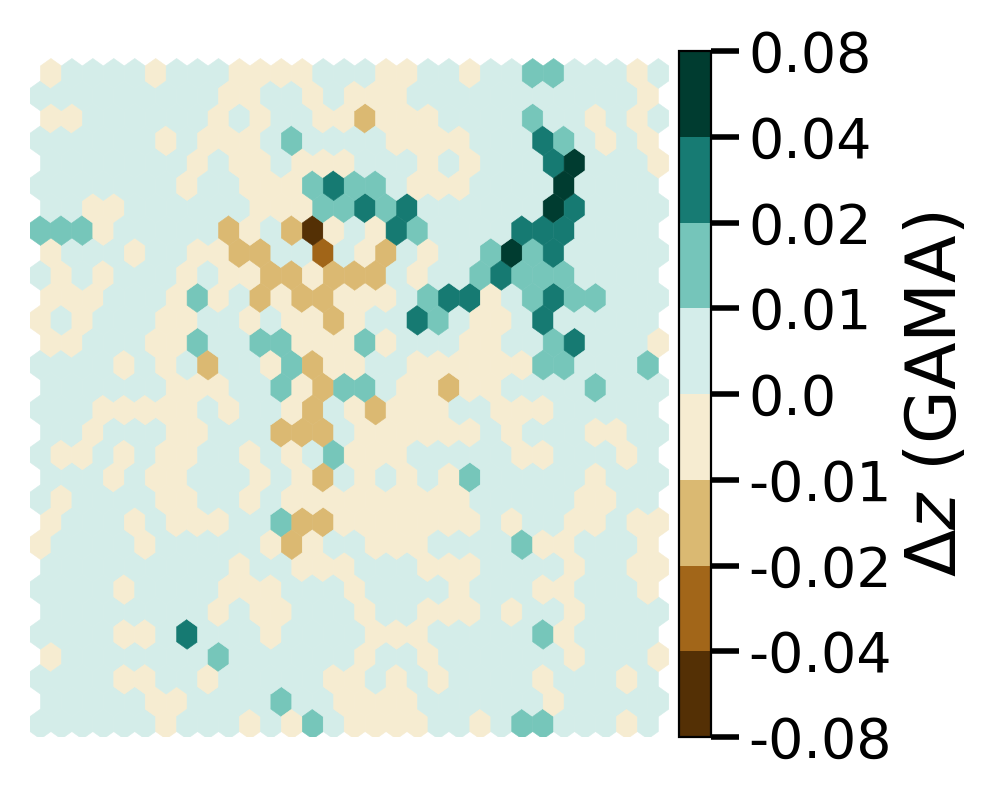}
\includegraphics[width=0.22\textwidth]{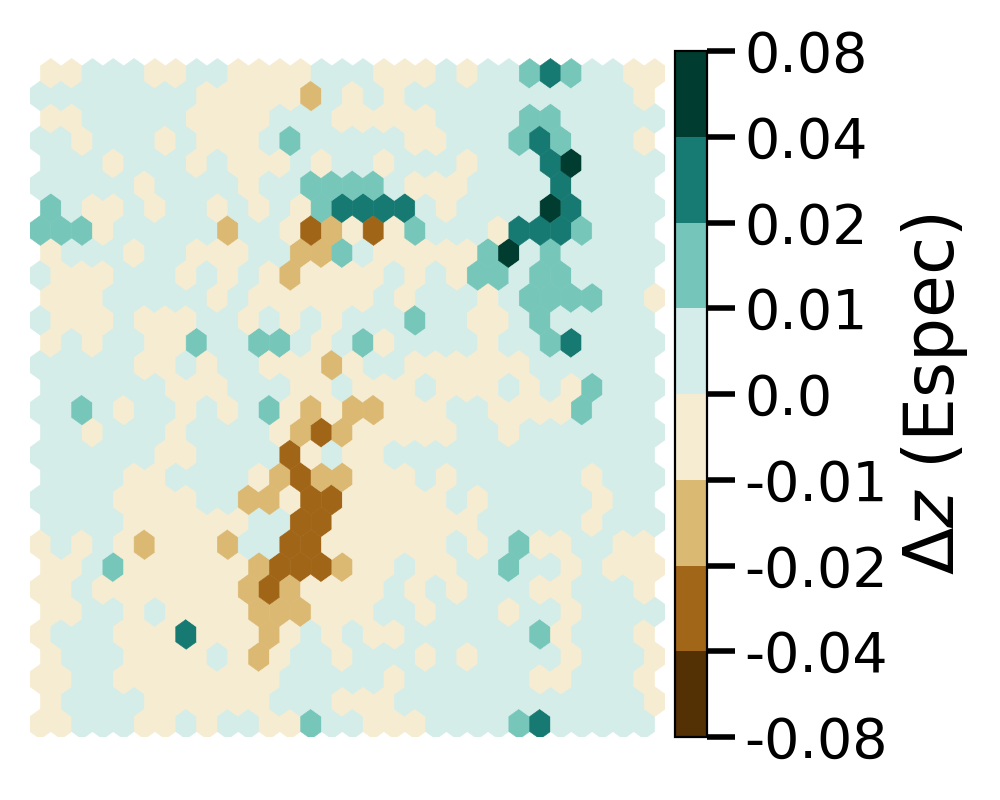}
\caption {The absolute difference $\Delta z \equiv \delta z / (1+z_\mathrm{spec})$ between the KiDS-Bright photometric and calibration spectroscopic redshifts, calculated in each SOM cell. The left and right panel shows  $\Delta z$ for GAMA and Espec, respectively.}
\label{Fig:zabs}
\end{figure}

Using the SOM projections discussed above, we can define per-cell $Dz$ and $\Delta z$ by replacing the respective $z_\mathrm{phot}$ and $z_\mathrm{spec}$ with their SOM-based averages. This is illustrated in Fig.~\ref{Fig:zrel} for $Dz_\mathrm{SOM}$ and in Fig.~\ref{Fig:zabs} for $\Delta z_\mathrm{SOM}\equiv (\langle z_\mathrm{phot} \rangle_\mathrm{cell} - \langle z_\mathrm{spec} \rangle_\mathrm{cell})/(1+\langle z_\mathrm{spec} \rangle_\mathrm{cell}) $. Each of the figures includes two panels: one for computing differences with respect to GAMA-only spectra (left), and one for computing differences with respect to the full Espec compilation (right). The first observation that we can make from these figures is that, in most of the cells, both relative and absolute photo-$z$ errors are small. For $Dz$, the majority are within 10\%, while typically $|\Delta z|<0.01$ except for a small fraction of cells. There are, however, some notable differences between the cases of GAMA-only spectra and Espec spectra. As only the former was used for photo-$z$ estimation, such deviations could indicate the cells where GAMA does not provide sufficiently robust redshift calibration. On the other hand, the overall consistency between the respective left- and right-hand panels suggests that GAMA does provide sufficiently good photo-$z$ calibration for most of the KiDS-Bright sample. 

Following the considerations above, we propose three possible ways of cleaning up the KiDS-Bright sample to improve its match with overlapping spectroscopic datasets. The first is a simple flux cut in the $r$-band, namely adopting $r_\mathrm{lim}$ lower than the original $r<20$ mag selection. The second is a threshold in $|Dz|$, while the third one assumes a cut in $|\Delta z|$. For a direct comparison, each of these cuts are implemented on the SOM, based on per-cell averages: that is, we remove from KiDS-Bright all the galaxies belonging to the cells where the respective cell-average is above the given threshold. This means, in particular, that the flux limit selection will not perform an exact cut at $r<r_\mathrm{lim}$, as each cell has some non-zero intrinsic scatter in $r$-band magnitudes. Furthermore, we define our thresholds in such a way as to remove approximately the same fraction of KiDS-Bright galaxies with each selection. We choose this fraction to be roughly 10\%, which leads to the following cuts applied on the SOM cells:
\begin{enumerate}
    \item[{[1]}] $\langle r \rangle < 19.83$ mag;  
   \item[{[2]}] $|Dz| < 5.97\%$;
    \item[{[3]}] $|\Delta z| < 0.01$;  
    \item[{[4]}] A combination of the above criteria: all the SOM cells that have $\langle r \rangle < 19.86$ mag AND $|Dz| < 10.05\%$ AND $|\Delta z| < 0.015$;
\end{enumerate}
where the values are calculated using per SOM cell averages and the particular cuts are applied separately one at a time. We also reiterate that the criteria  [2]-[4] could use either GAMA-only or Espec to compute $Dz$ and $\Delta z$. To keep the fraction $\sim10\%$, the above constraints changes slightly when Espec is used instead of GAMA. In option [4], each of the particular selections removes roughly 4-5\% of galaxies, but their combination gives about 10\%, as in the separate cuts [1]-[3]. This is due to considerable overlap between removed SOM cells, especially between the $Dz$ and $\Delta z$ based selections. We illustrate this in the Appendix Fig.~\ref{Fig:somcells20} for a combination of 20\% galaxies removed.

\begin{figure}
\centering
\includegraphics[width=0.49\textwidth]{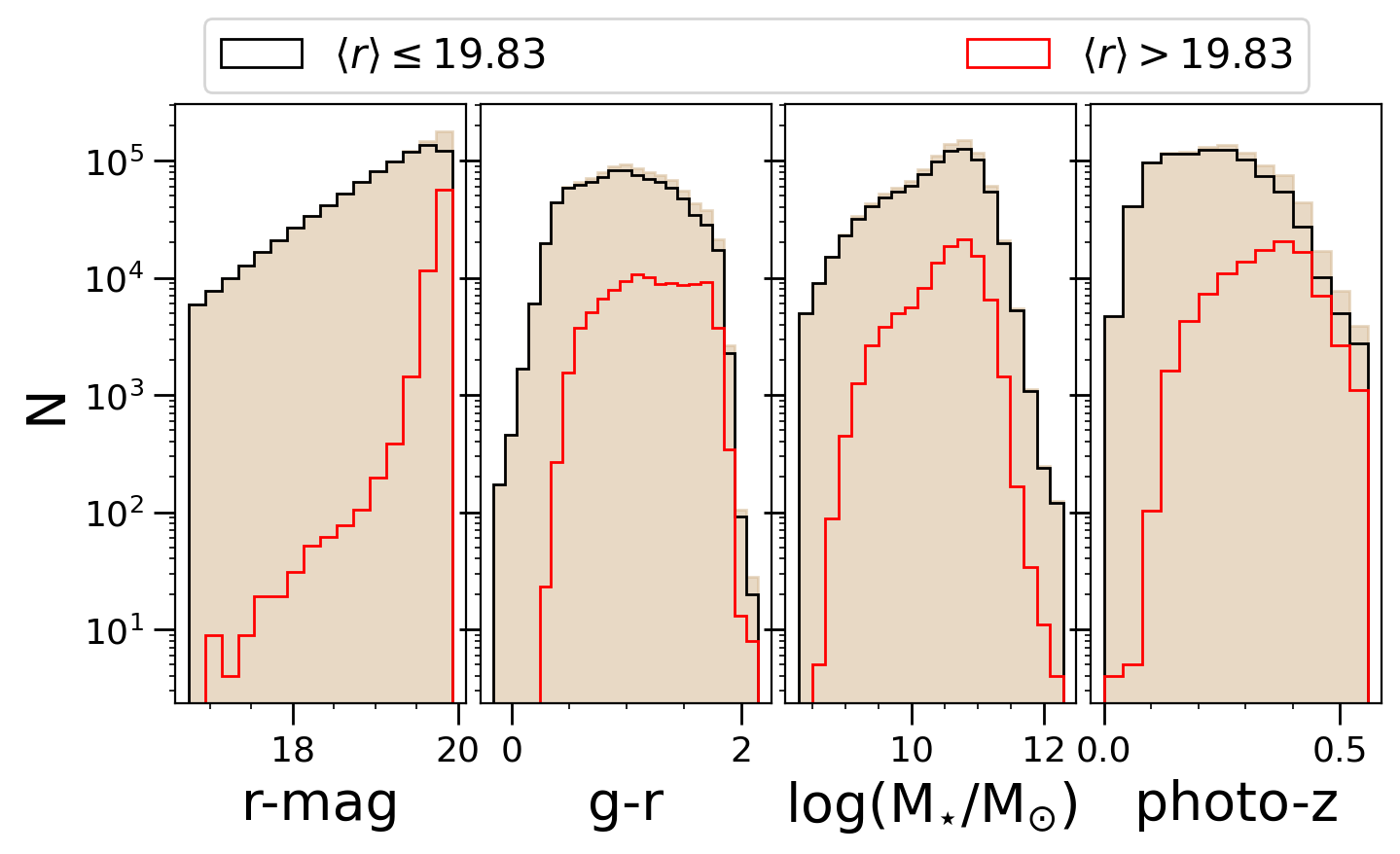}
\caption{Distribution of $r$ magnitude, $g-r$ color, stellar mass, and photometric redshifts for KiDS-Bright galaxies. Filled tan: before any cuts; black: keeping the SOM cells with $\langle r \rangle <19.83$; red: removed objects from cells with  $\langle r \rangle > 19.83$ (about 10.4\% of all).}
\label{Fig:rcut_hist}
\end{figure}

\begin{figure}  
\centering
\includegraphics[width=0.47\textwidth]{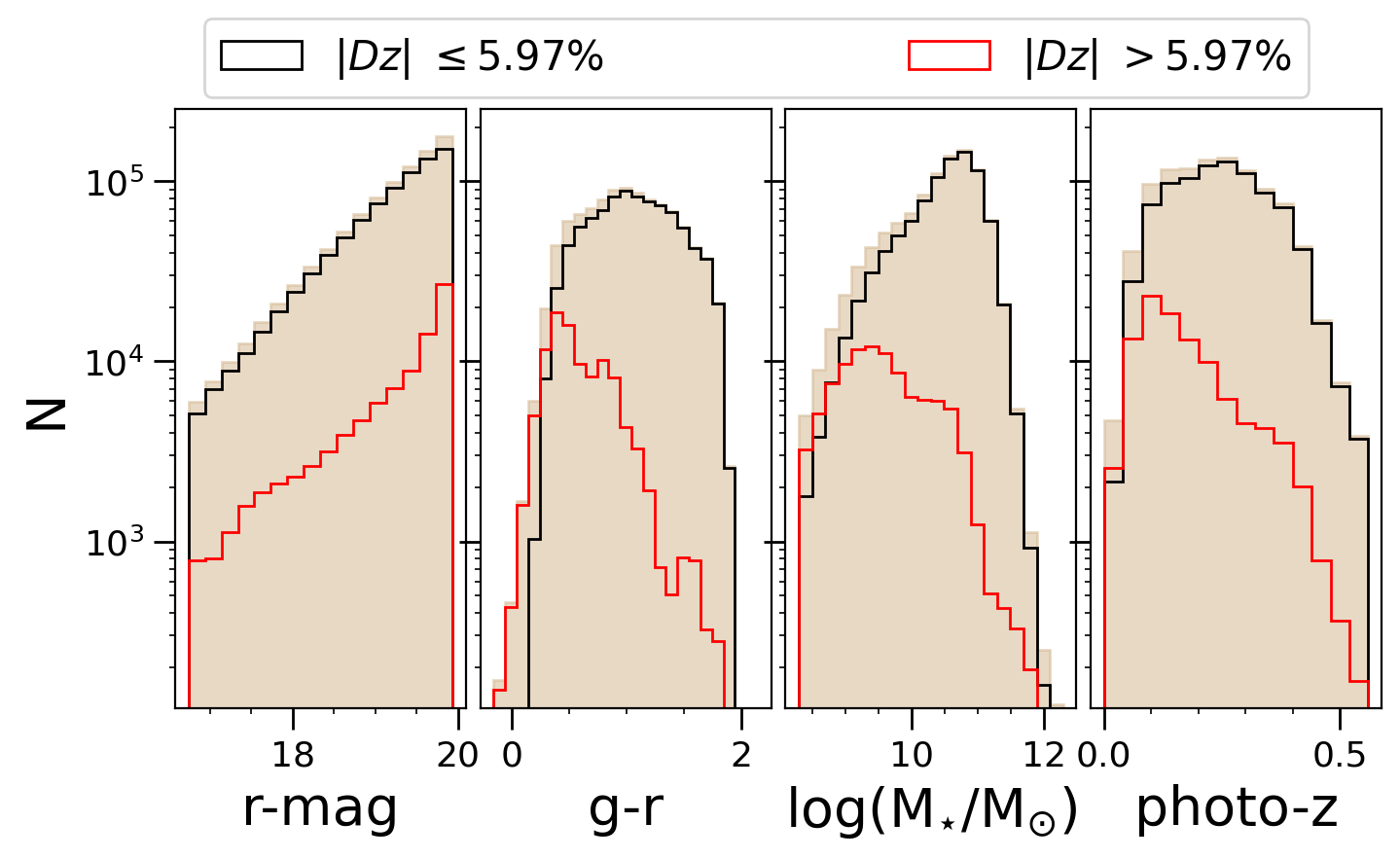}
\includegraphics[width=0.47\textwidth]{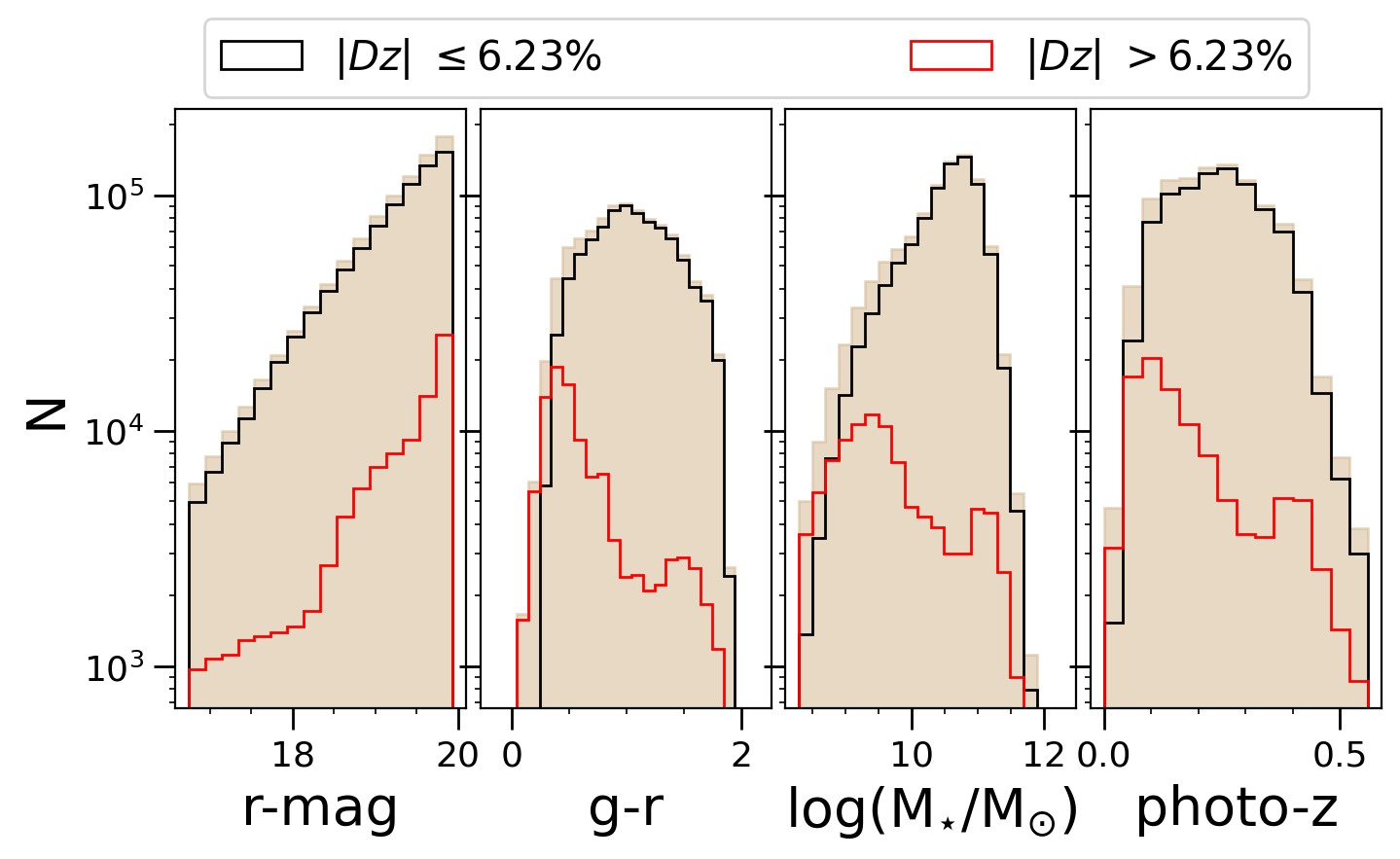}
\caption{{\it Upper panels:} Same as Fig.~\ref{Fig:rcut_hist}, but the black lines indicate galaxies with SOM cells where $|D z| = |z_\mathrm{phot}-z_\mathrm{spec}|/z_\mathrm{spec}<5.97\%$ (relative redshift error), while red illustrates objects ($\sim 10\%$) from removed cells. GAMA is used as the source of spectroscopic redshifts. {\it Lower panels:} Spec-$z$ taken from Espec compilation  ($|D z| < 6.23\%$). }
\label{Fig:zrel_hist}
\end{figure}

\begin{figure}  
\centering
\includegraphics[width=0.49\textwidth]{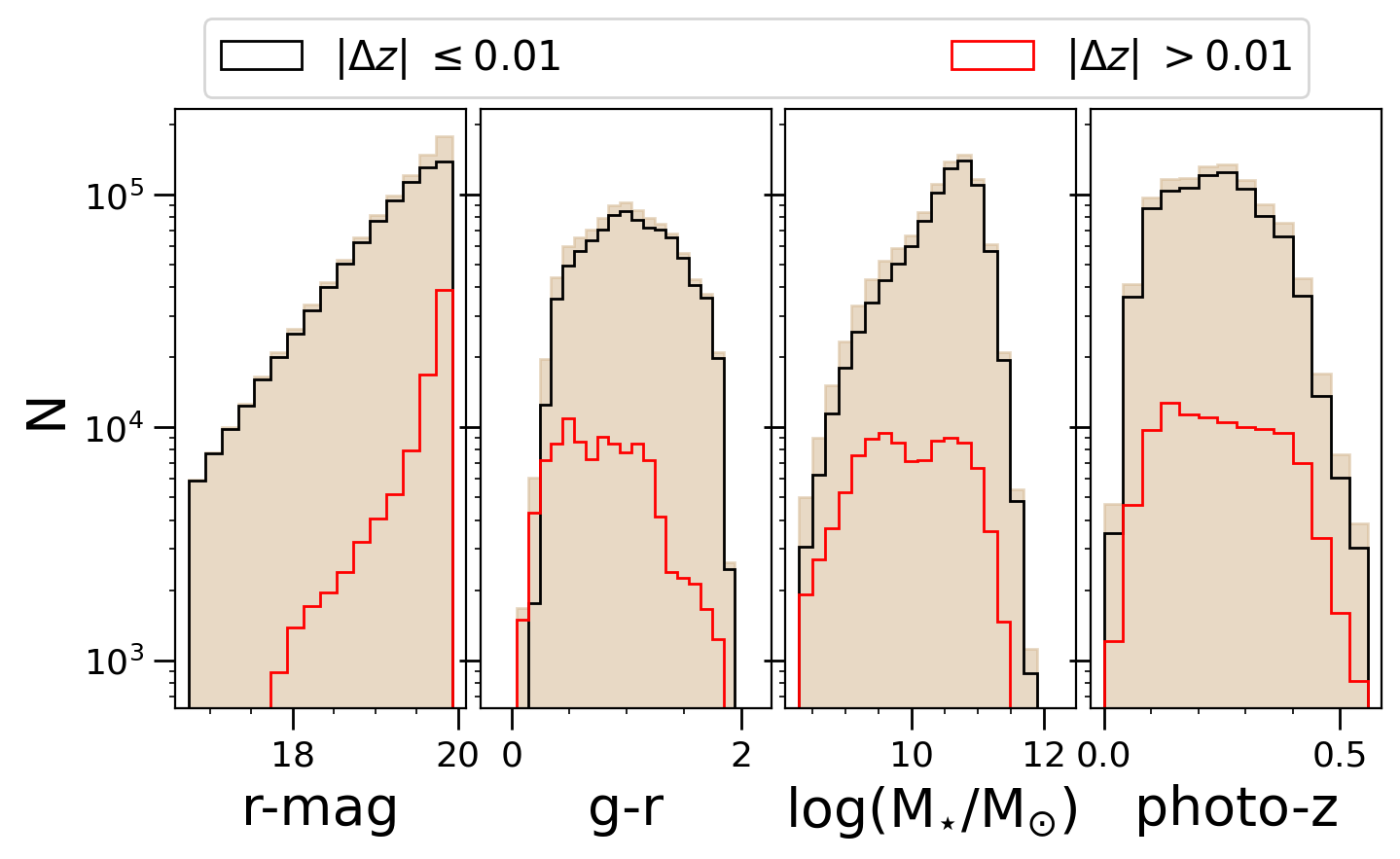}
\includegraphics[width=0.49\textwidth]{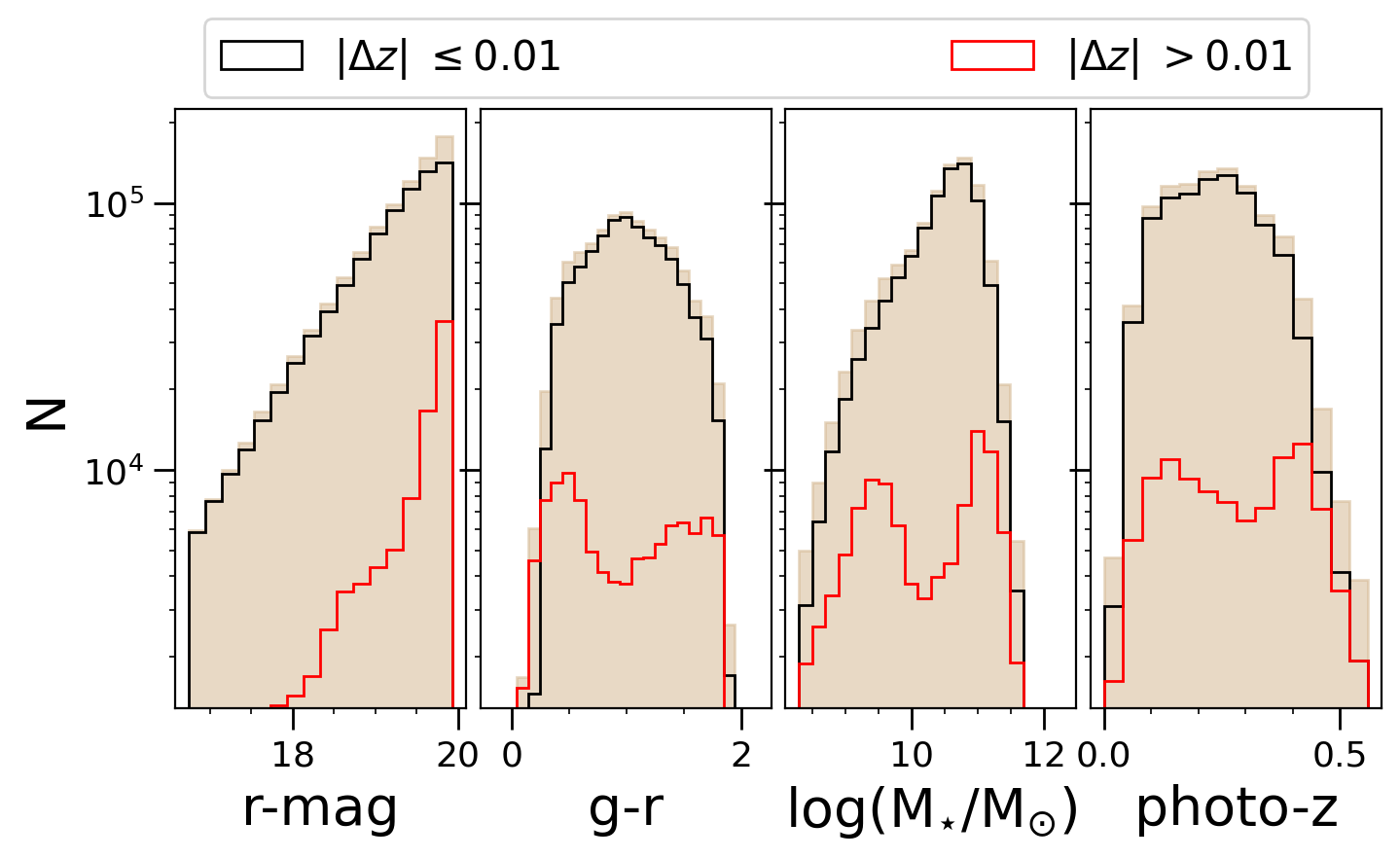}
\caption{Same as Fig.~\ref{Fig:rcut_hist} but the black
lines indicate galaxies with SOM cells where $|\Delta z| = |z_\mathrm{phot}-z_\mathrm{spec}|/(1+z_\mathrm{spec})<0.01$ (absolute redshift error), while red illustrates objects ($\sim 10\%$) from removed cells. {\it Upper panels:} GAMA is used as the source of spectroscopic redshifts. {\it Lower panels:} Spec-$z$ taken from Espec compilation.}
\label{Fig:zabs_hist}
\end{figure}

\begin{figure}
\centering
\includegraphics[width=0.49\textwidth]{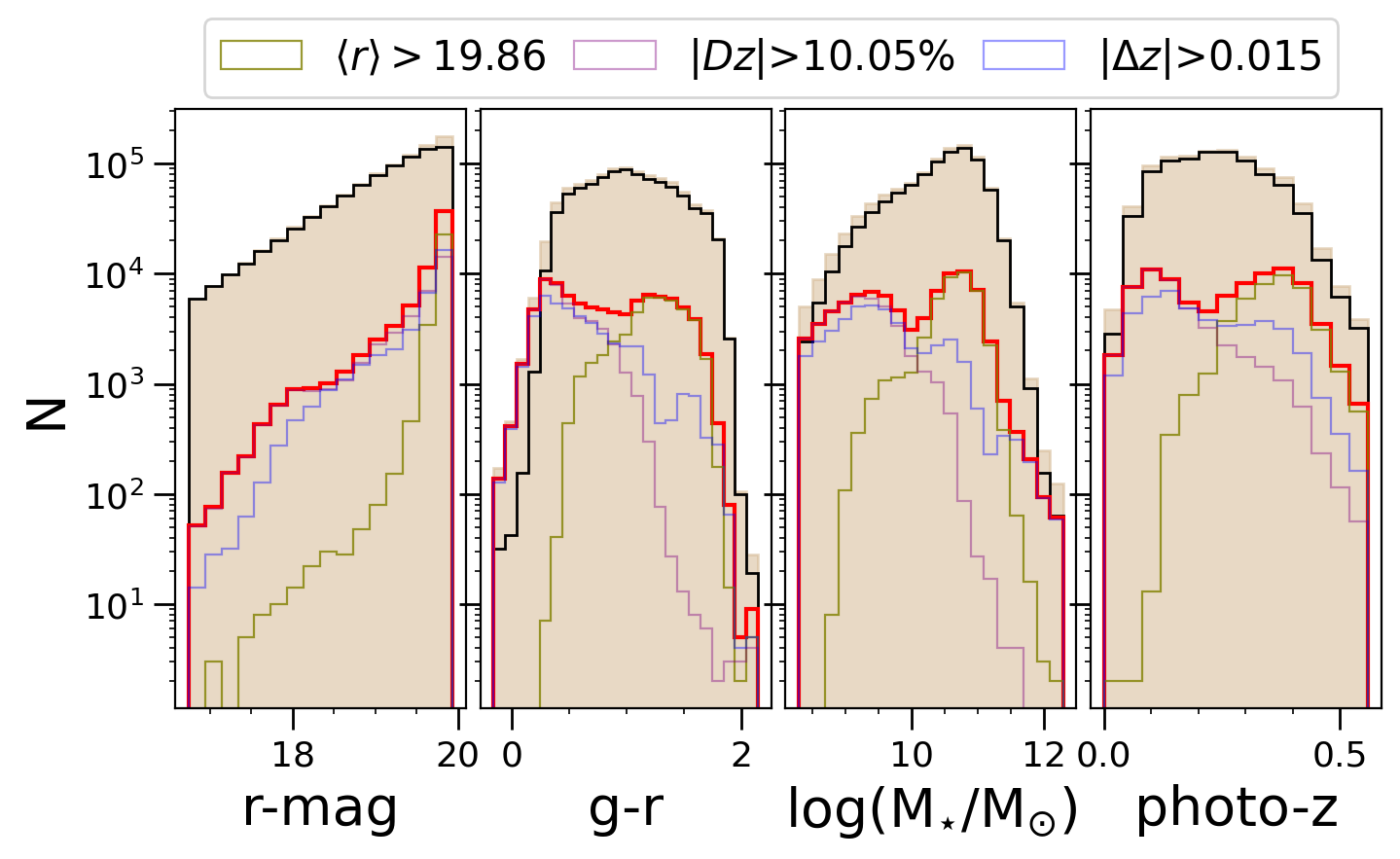}
\includegraphics[width=0.49\textwidth]{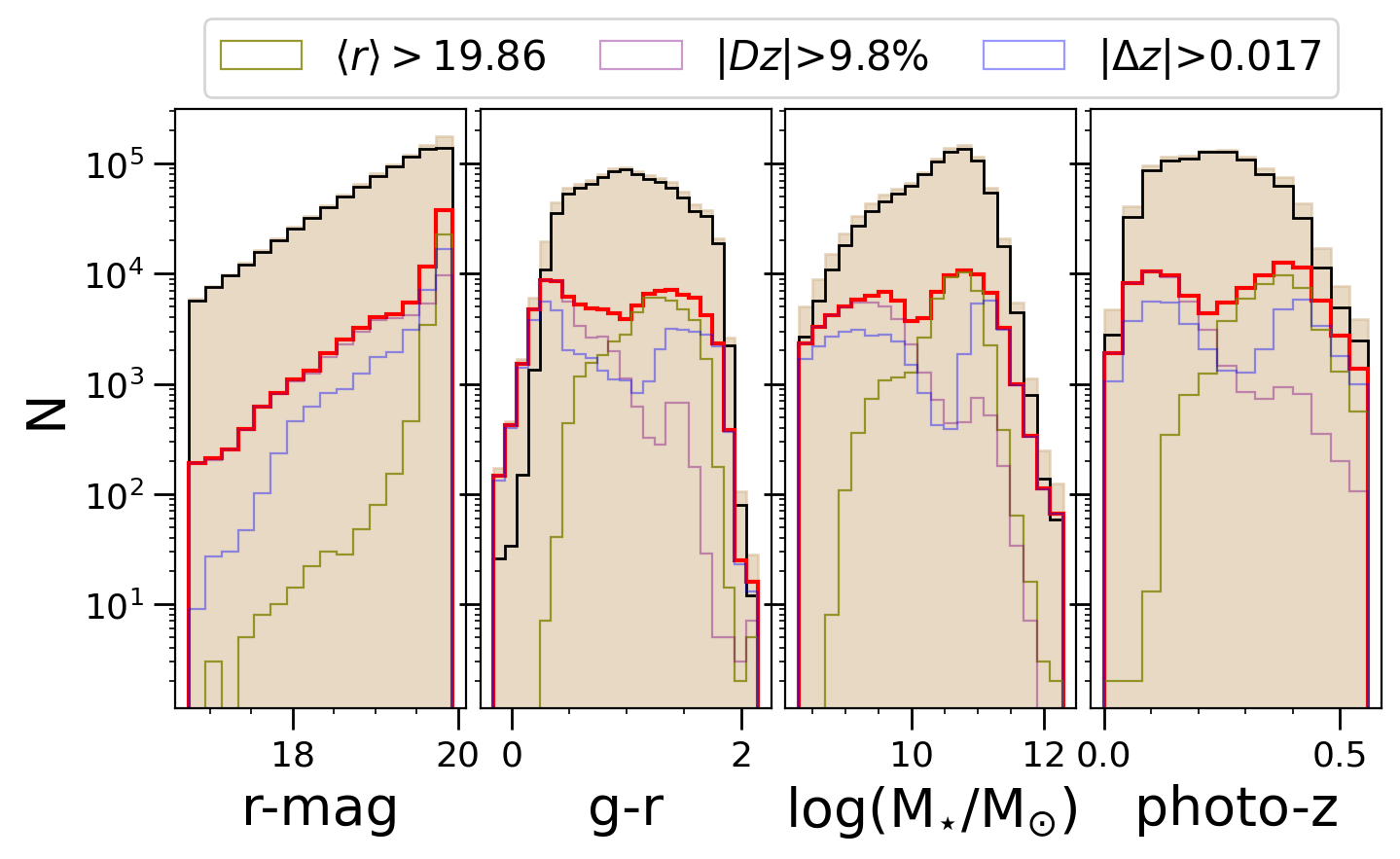}
\caption{Same as Fig.~\ref{Fig:rcut_hist}, but the black lines indicate galaxies with SOM cells kept after the joint condition on $\langle r\rangle$ mag, $Dz$ and $\Delta z$ is applied (see text for details), while red illustrates objects from all the removed cells. Here we additionally show the distribution for each individual removal from the combination: for $r$ mag in green, $D z$-based in purple, and for $\Delta z$ in blue. The combined criteria remove $\sim10\%$ of the objects. {\it Upper panels:} GAMA is used as the source of spectroscopic redshifts. {\it Lower panels:} Spec-$z$ taken from Espec compilation.}
\label{Fig:joint}
\end{figure}

In Figs.~\ref{Fig:rcut_hist}-\ref{Fig:zabs_hist} we show the distributions of the same galaxy properties as analyzed earlier in Sect.~\ref{s4:analysis}, but now comparing three cases in each panel: before any cuts to the KiDS-Bright dataset (filled tan), after the cuts (black), and additionally what is removed by the cuts (red). Fig.~\ref{Fig:rcut_hist} applies to the SOM cells with the faintest sources being removed (item [1] above), Fig.~\ref{Fig:zrel_hist} depicts the case where the cells with the largest relative photo-$z$ error $Dz$ are taken out (item [2]), while Fig.~\ref{Fig:zabs_hist} illustrates the removal of the biggest absolute photo-$z$ errors ($\Delta z$, item [3]). Finally, Fig.~\ref{Fig:joint} illustrates the joint condition [4] discussed above. For item [4], we also show the histograms for the objects discarded by the individual cuts that are combined, each of which culls about 4-5\% of KiDS-Bright galaxies. Figures \ref{Fig:zrel_hist}-\ref{Fig:joint} are composed of two sets of panels each, those in the top based on using GAMA only as spec-$z$ calibration, while the bottom ones employ the entire Espec compilation for the cuts. We can summarize the results as follows:
\begin{enumerate}
    \item Removing the SOM cells at the faint end affects mostly galaxies that are redder, have higher stellar masses, and are at higher redshifts than the rest of the KiDS-Bright sample. This is expected for a flux-limited dataset, where the (observationally) faint end is dominated by intrinsically bright, red, massive galaxies, which  are typically observed at higher redshifts than the sample mean.
    \item Discarding the cells with the largest relative photo-$z$ residuals, $Dz$, affects galaxies at a range of magnitudes, but mainly at the faint end. Unlike in case [1], however, the removed objects are mostly blue, of low stellar mass, and at lower redshifts. This is consistent with the fact that the relative photo-$z$ errors become very large at low redshift and additionally indicates that $Dz$ is typically larger for low-mass, blue, and intrinsically faint galaxies. 
    \item Cleaning up the KiDS-Bright sample of the SOM cells that have the largest absolute photo-$z$ errors, $\Delta z$, also gives quite different outcomes than cases [1] and [2]. While it is still preferably galaxies at the fainter (observed) magnitudes that are affected, and practically all the bluest objects are discarded this way, the distributions of the $g-r$ color, stellar mass, and photo-$z$ of the removed sources otherwise sample these parameters much more uniformly. In other words, requiring a fixed cut in absolute photo-$z$ error seems to remove galaxies equally as a function of color, stellar mass, and photo-$z$. It is however important to note that this is not equivalent to a sparse sampling, as the distributions of the removed objects are very different from those of the entire input population.
    \item Finally, combining the three above types of cuts into one condition, but keeping $\sim90\%$ of galaxies, again preferentially removes the objects at the faint end, but their distribution in other properties is also more uniform than in cases [1]-[3].
\end{enumerate}
Moreover, we conclude that none of these cuts heavily influences the completeness curve presented in Fig.~\ref{Fig:ratio_rzug}.

By comparing the upper and lower panels of Figs.~\ref{Fig:zrel_hist}-\ref{Fig:joint}, we can see some differences between the galaxies affected by the cuts when using GAMA versus Espec as calibration. This can indicate situations where GAMA does not provide sufficient calibration for KiDS-Bright color space and additional data, such as from Espec, should be used to purify the photometric sample. However, as discussed below, interchanging between the two spectroscopic datasets for cleanup has very little influence on the resulting photo-$z$ statistics.

\begin{table*}[ht]
\begin{center}
\caption{Statistics of photometric redshift performance for the KiDS-Bright sample cross-matched with GAMA-eq. Numbers in columns (4)-(7) refer to subsamples kept after applying the cuts in the first column.
\label{Tab: photo-zstats}}
\begin{tabular}{lrrrrrr}
\hline
\centering { } {\textbf{Selection} } & { \textbf{Number of} } & \multicolumn{1}{c}{ \textbf{Mean} } & \multicolumn{1}{c}{ \textbf{Mean of} } & \multicolumn{1}{c}{ \textbf{Mean of} } & \multicolumn{1}{c}{\textbf{St. dev.\ of}} & \multicolumn{1}{c}{\textbf{SMAD of}} \\

{} & \multicolumn{1}{c}{ \textbf{galaxies} } & \multicolumn{1}{c}{ \textbf{redshift} } & \multicolumn{1}{c}{$\delta z = z_\mathrm{phot}-z_\mathrm{spec}$ } &  \multicolumn{1}{c}{$\delta z/(1+z_\mathrm{spec})$ } &  \multicolumn{1}{c}{$\delta z/(1+z_\mathrm{spec})$ } &  \multicolumn{1}{c}{$\delta z/(1+z_\mathrm{spec})$} \\ 

\hline                                                                            No cut & $10.00\times10^5$ &  0.242 &   $4.7\times10^{-4}$ &  $9.1\times10^{-4}$ & 0.0232 & 0.0178 \\
$\langle r \rangle < 19.83$ &  $8.96\times10^5$ &  0.230 &   $4.8\times10^{-4}$ &  $9.1\times10^{-4}$ & 0.0230 & 0.0176 \\
$|Dz|^{a}<5.97\%$ (GAMA-based) &  $8.97\times10^5$ &  0.250 &  $-1.8\times10^{-4}$ &  $3.2\times10^{-4}$ & 0.0219 & 0.0174 \\
$|Dz| < 6.23\%$ (Espec-based) &  $8.98\times10^5$ &  0.248 &  $-0.7\times10^{-4}$ &  $4.1\times10^{-4}$ & 0.0217 & 0.0173 \\
$\Delta z^{b} < 0.01$ (GAMA-based) &  $8.96\times10^5$ &  0.240 &  $-1.0\times10^{-4}$ &  $4.0\times10^{-4}$ & 0.0218 & 0.0174 \\
$\Delta z < 0.01$ (Espec-based) &  $8.96\times10^5$ &  0.237 &  $-0.2\times10^{-4}$ &  $4.6\times10^{-4}$ & 0.0218 & 0.0174 \\
Combined (GAMA-based)$^1$ &  $9.11\times10^5$ &  0.240 &   $0.4\times10^{-4}$ &  $5.2\times10^{-4}$ & 0.0220 & 0.0175 \\
Combined (Espec-based)$^2$ &  $9.01\times10^5$ &  0.238 &  
  $0.4\times10^{-4}$ & $5.1\times10^{-4}$ & 0.0219 & 0.0174 \\
\hline
\end{tabular}
\begin{tablenotes}
The cuts are based on the SOM cell averages. \\
$^a$ $Dz\equiv (\langle z_\mathrm{phot} \rangle_\mathrm{cell} - \langle z_\mathrm{spec} \rangle_\mathrm{cell})/\langle z_\mathrm{spec} \rangle_\mathrm{cell} $ \\ 
$^b$ $\Delta z\equiv (\langle z_\mathrm{phot} \rangle_\mathrm{cell} - \langle z_\mathrm{spec} \rangle_\mathrm{cell})/(1+\langle z_\mathrm{spec} \rangle_\mathrm{cell}) $ \\
$^1$ $\langle r \rangle  < 19.86$ mag \&  $|Dz|$ $< 10.05\%$ \& $\Delta z < 0.015$. \\
$^2$ $\langle r \rangle < 19.86$ mag \&  $|Dz|$ $< 9.8\%$ \& $\Delta z < 0.017$
\end{tablenotes}
\end{center}
\end{table*}

The effect of the KiDS-Bright sample cleanup on its photo-$z$ properties is quantified in Table \ref{Tab: photo-zstats}. As expected, the average redshift of the sample decreases if a brighter flux cut is adopted, while it goes up if the galaxies with the largest relative photo-$z$ error $Dz$ (Eq.~\ref{eq:rel}) are discarded. It remains largely unchanged if the cleanup is based on the largest absolute errors $\Delta z$. The photo-z statistics are calculated for the  KiDS-Bright sample cross-matched with GAMA-eq. The reduction in the mean photo-$z$ residuals (Eq.~\ref{eq:abs}, columns 4-5) is by a factor of a few or even more (in absolute terms) for redshift-based cuts while interestingly they stay almost intact for a more aggressive magnitude cut. Furthermore, each of the analyzed cuts lowers the scatter in photo-$z$ errors, both in terms of standard deviation and SMAD, when compared to the original sample case: the former is reduced by $\sim 6\%$ (when implementing cleanup in $Dz$ or $\Delta z$), whereas the latter decreases by $\sim3\%$ (cut in $Dz$ based on the Espec calibration). It is worth noting that a larger relative reduction in SD than in SMAD means that the photo-$z$ residuals become more Gaussian after our proposed cuts, due to a reduction in outliers. Still, the fact that SD is still much larger than SMAD indicates that there are still considerable non-Gaussian wings and/or outlier populations in the photo-$z$ error distribution (see, e.g., the discussion in \citetalias{Bilicki2021}).

Finally, we note that the same general observations hold when we use more conservative selection cuts and remove twice as many galaxies (i.e., 20\%) from the sample; see Appendix \ref{app:twenty-percent}.
In fact, one would need to remove the faintest half of the dataset (cutting at $r\lesssim19.4$ mag) to obtain only 10\% reduction in photo-$z$ scatter, as we shortly discuss in Appendix \ref{app:percentages}. Similarly, reducing the mean residuals to $\sim0$ would require rejecting over 30\% of the KiDS-Bright galaxies. Such aggressive cleanups would considerably affect the usefulness of the dataset, especially for GGL science. Indeed, as shown in\ \citetalias{Bilicki2021}, for example, for this sample, it is the actual number of galaxies used as lenses that drives the uncertainties of the GGL measurements, rather than the photo-$z$ errors. In other words, to improve the scientific usefulness of the KiDS-Bright dataset, it would be more desirable to increase its number density, rather than to reduce it.

\section{Summary and conclusions}
\label{Sect:conc}
In this paper, we employed self-organizing maps (SOMs) to analyze how well the photometric KiDS-Bright galaxy sample, flux-limited at $r<20$ mag, is matched in the color space with the available overlapping spectroscopic data samples. This analysis is particularly important because applications of the KiDS-Bright sample often require robust photometric redshift estimates. Therefore, it is crucial to validate the performance of the current ANNz2-based photo-$z$ \citepalias{Bilicki2021} across various galaxy types and brightness ranges.

We first focused on the GAMA-Equatorial (GAMA-Eq) spec-$z$ sample, which was used by \citetalias{Bilicki2021} as a training dataset in ANNz2 to derive KiDS-Bright photo-$z$. GAMA-Eq was originally designed to be highly complete spectroscopically up to $r<19.8$ in Petrosian magnitude, as derived from SDSS \citep{Baldry2010}. Its eventual completeness was, however, established to be 98\% at $r<19.6$ using  KiDS imaging \citep{Driver2022GAMA}, which is 0.4 mags brighter than the KiDS-Bright flux limit. Nonetheless, as GAMA includes galaxies with spec-$z$ measurements fainter than this limit (typically filler targets), \citetalias{Bilicki2021} was able to train KiDS-Bright photo-$z$ beyond the GAMA completeness limit. Our SOM analysis shows that while GAMA-Eq completeness, relative to KiDS-Bright, quickly declines for magnitudes fainter than $r\sim19.5$, there is no significant trend in relative incompleteness with multiwavelength colors or stellar mass. This suggests that, despite missing many faint-end galaxies, GAMA-Eq does provide a good representation of the KiDS-Bright galaxy population over the color space. 

Having compared the KiDS-Bright and GAMA in the color space using SOM, we next proposed a possible approach to further enhance the photo-$z$ quality of the former, by cleaning the target sample to force a better match with the overlapping spectroscopy. Here, we extend our study beyond GAMA, by adding spec-$z$ from several other datasets overlapping with KiDS, as well as with external calibration fields dubbed "KiDZ." 

By comparing KiDS color-space projections with those from spectroscopic samples, we identified the SOM cells where KiDS-Bright photo-$z$ differed the most from the true redshifts, both in absolute and relative terms. This allowed us to propose four possible criteria for KiDS-Bright cleanup: [1] based on reduced magnitude limit; [2] removing the cells with the largest $|D z| \equiv |z_\mathrm{phot}-z_\mathrm{spec}|/z_\mathrm{spec}$; [3] cutting out those cells where $|\Delta z| \equiv |z_\mathrm{phot}-z_\mathrm{spec}|/(1+z_\mathrm{spec})$ is the biggest; and finally [4] a combination of the three previous options. For each criterion, we chose to remove SOM cells such that roughly 10\% of galaxies are discarded from the KiDS-Bright sample. This fraction can of course be adjusted based on the specific scientific goals of an analysis, balancing between the statistical power of the dataset (in terms of the total number of galaxies retained) and the accuracy and precision of the photo-$z$. In the appendix, we discuss an alternative scenario of removing about 20\% of input galaxies.  Finally, we note that each of the criteria [1]-[4] affect different galaxy types, in terms of their magnitudes, colors, stellar masses, and/or photo-$z$. 

We conclude that none of the cleanups that we analyzed led to significant improvement in the overall photo-$z$ quality of the KiDS-Bright sample. The SMAD values are never reduced to below 3\%, even after 10\% reduction in the sample. On the one hand, this may indicate that the current KiDS-Bright photo-$z$ are already close to being optimal for the set of passbands that KiDS+VIKING provide. 

Our findings presented here can be further scrutinized in the near future by supplementing the spec-$z$ data we used here with additional redshifts. This could be done by either selecting galaxies from the most under-represented cells for targeted spectroscopic follow-up (similarly as done in the C3R2 projects, \citealt{Masters2017C3R2}) or by gradually supplementing these cells with spec-$z$ coming from large surveys. In this latter respect, DESI looks particularly promising, as it includes a Bright Galaxy Survey \citep{Hahn2023}, in which the faint sub-survey is obtaining redshifts for sources up to $r<20.175$. In our work, we include DESI Early Data Release galaxies \citep{DESI2023}, but their overlap with our photometric sample is evidently too small to considerably influence our findings. We hope to be able to further test our approach with the forthcoming DESI Data Release 1 from the main survey. On a longer term, the 4-meter Multi-Object Spectroscopic Telescope \citep[4MOST,][]{Jong2019} will undertake a number of surveys that should provide a good match to KiDS both in sky coverage and color space.

In this work, we focused on the KiDS DR4 Bright Sample, but our framework can be adapted to any galaxy dataset where such an analysis is relevant. In particular, we plan to employ the selection of foreground galaxies from the final data of KiDS DR5 \citep{Wright2024}, where a new version of the KiDS Bright Sample will be designed as the main dataset for GGL and clustering in KiDS 3$\times$2pt analyses.
Furthermore, our methodology will be relevant for forthcoming deep wide-area surveys such as from Euclid or LSST. While a fraction of galaxies detected by these surveys will have been covered by spectroscopy from DESI, 4MOST, or Euclid itself, on their full area the most complete selection of foreground galaxies for clustering and galaxy-galaxy lensing may still be possible only from photometry. It will be then natural to apply similar approaches to photo-z derivation for, for instance, KiDS-Bright (i.e., by training some ML models), which will equally need quantification of the match between spectroscopic training sets and target photometrically selected samples. Our framework presented here could be practically directly employed once these new data become available. We can then follow up with an appropriate clean-up of the photometric datasets and observational proposals to obtain spectroscopy for the least complete SOM cells.

\begin{acknowledgements}
We thank our anonymous referee for their positive feedback and useful comments. PJ and MB are supported by the Polish National Science Center through grant no. 2020/38/E/ST9/00395. MB and WAH are supported by the Polish National Science Center through grants no. 2018/30/E/ST9/00698, 2018/31/G/ST9/03388 and 2020/39/B/ST9/03494. AHW is supported by the Deutsches Zentrum für Luft- und Raumfahrt (DLR), made possible by the Bundesministerium f\"{u}r Wirtschaft und Klimaschutz, and acknowledges funding from the German Science Foundation DFG, via the Collaborative Research Center SFB1491 "Cosmic Interacting Matters - From Source to Signal". AD acknowledged support from ERC Consolidator Grant (No. 770935). This publication is part of the project ``A rising tide: Galaxy intrinsic alignments as a new probe of cosmology and galaxy evolution'' (with project number VI.Vidi.203.011) of the Talent programme Vidi which is (partly) financed by the Dutch Research Council (NWO). This work is also part of the Delta ITP consortium, a program of the Netherlands Organisation for Scientific Research (NWO) that is funded by the Dutch Ministry of Education, Culture and Science (OCW). CH acknowledges support from the European Research Council under grant number 647112, from the Max Planck Society and the Alexander von Humboldt Foundation in the framework of the Max Planck-Humboldt Research Award endowed by the Federal Ministry of Education and Research, and the UK Science and Technology Facilities Council (STFC) under grant ST/V000594/1. HHi is supported by a DFG Heisenberg grant (Hi 1495/5-1), the DFG Collaborative Research Center SFB1491, as well as an ERC Consolidator Grant (No. 770935). KK acknowledges support from the Royal Society and Imperial College. CM acknowledges support under the grant number PID2021-128338NB-I00 from the Spanish Ministry of Science, and from the European Research Council under grant agreement No. 770935. SJN is supported by the US National Science Foundation (NSF) through grant AST-2108402. JLvdB is supported by an ERC Consolidator Grant (No. 770935). ZY acknowledges support from the Max Planck Society and the Alexander von Humboldt Foundation in the framework of the Max Planck-Humboldt Research Award endowed by the Federal Ministry of Education and Research (Germany). MY acknowledges funding from the European Research Council (ERC) under the European Union’s Horizon 2020 research and innovation program (Grant Agreement No. 101053992). For the purpose of open access, the authors have applied a Creative Commons Attribution (CC BY) licence to any Author Accepted Manuscript version arising from this submission.

This work has made use of \textsc{TOPCAT} \citep{TOPCAT} software, as well as of \textsc{python} (\url{www.python.org}), including the packages \textsc{NumPy} \citep{NumPy}, \textsc{SciPy} \citep{SciPy}, \textsc{AstroPy} \citep{astropy}, \textsc{Pandas} \citep{pandas}, \textsc{Seaborn} \citep{seaborn} and \textsc{Matplotlib} \citep{Matplotlib}.\\

\textit{Author contributions:} All authors contributed to the development and writing of this paper. The authorship list is given in two groups: the lead authors (PJ, MB, WAH, AHW), followed by an alphabetical group which includes those who have either made a significant contribution to the data products or to the scientific analysis.

\end{acknowledgements}

\bibliography{ms_PJ}{}
\bibliographystyle{aa}
\appendix

\section{Removing 20 percent of the KiDS-Bright sample}
\label{app:twenty-percent}

\begin{figure}[h!]
\centering
\includegraphics[width=0.49\textwidth]{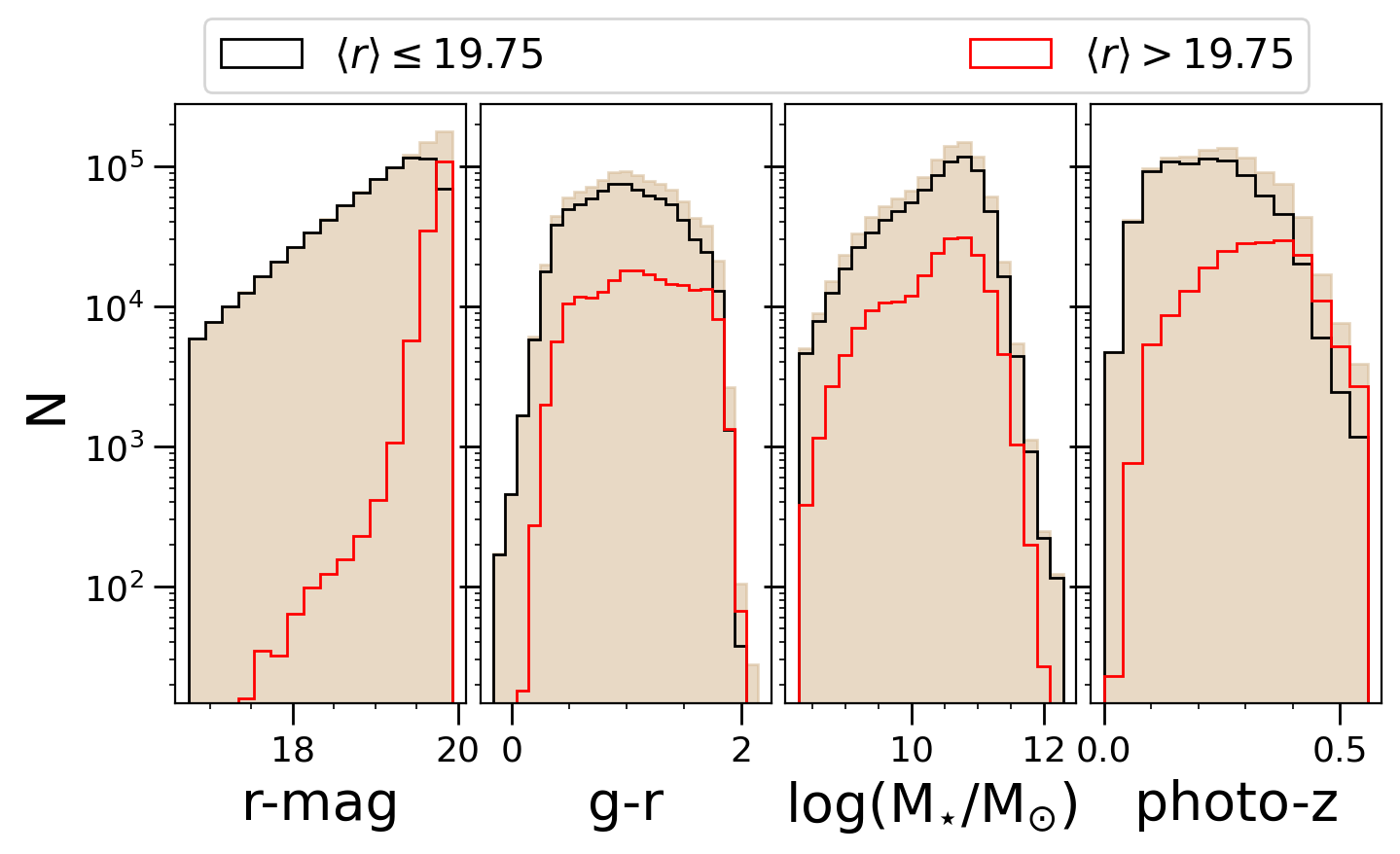}
\caption{Same as Fig.~\ref{Fig:rcut_hist}, but here the black [red] histogram shows the galaxies from the cells with the mean $r$ magnitude to be less [more] than  19.75, removing $\sim 20\%$ galaxies.} 
\label{Fig:rcut_hist20}
\end{figure}

\begin{figure}[h!]      
\centering
\includegraphics[width=0.49\textwidth]{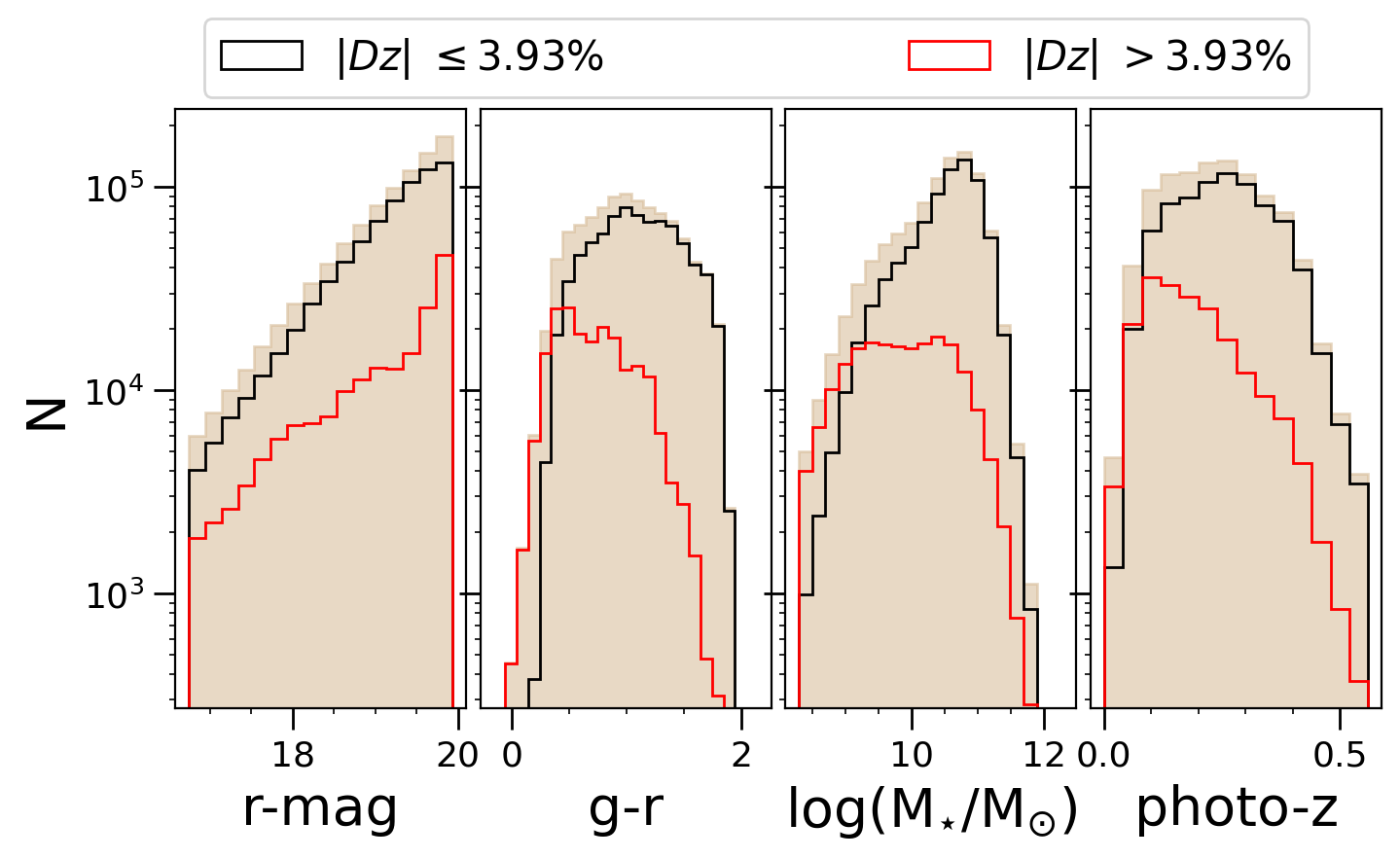}
\includegraphics[width=0.49\textwidth]{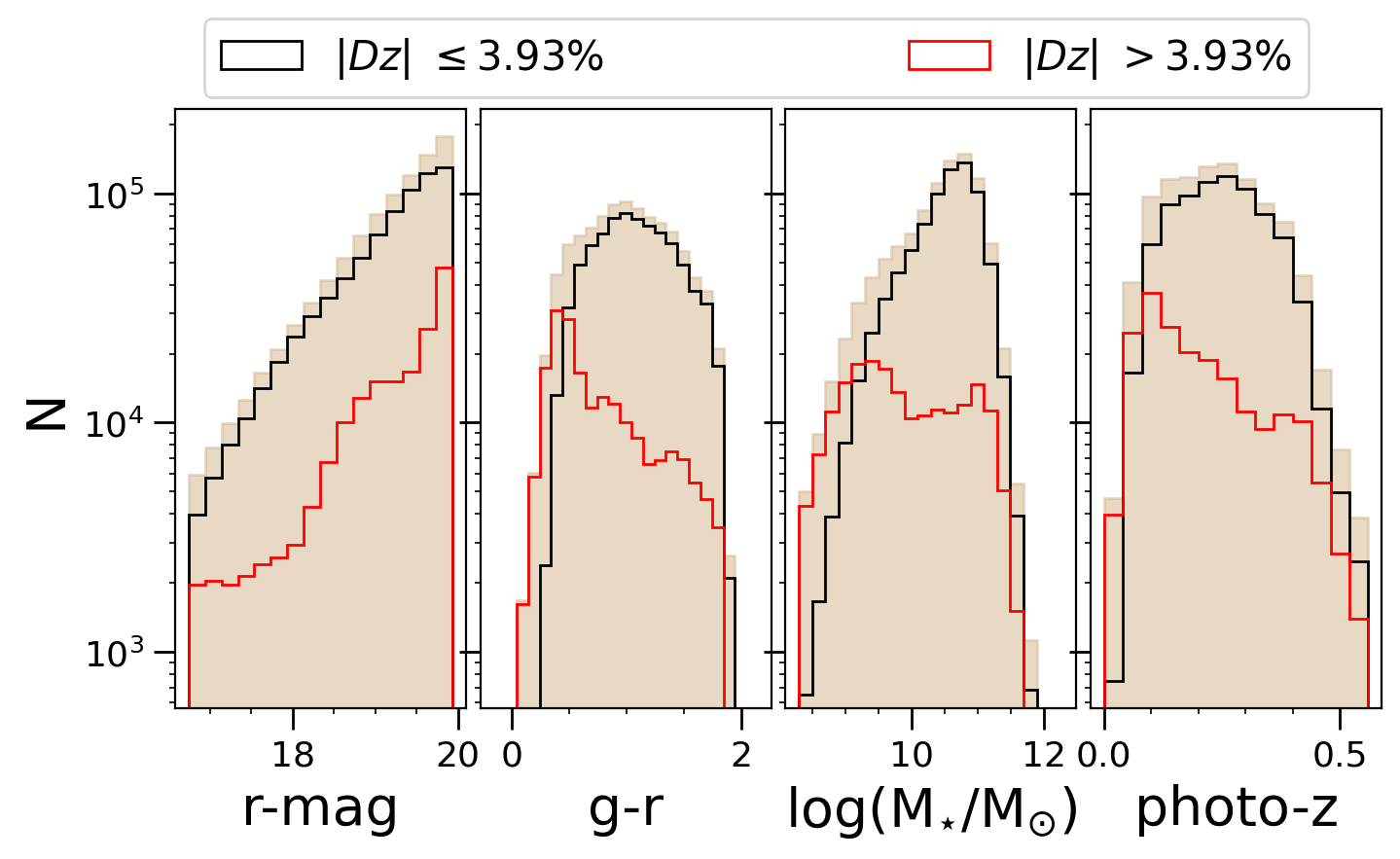}
\caption{Same as Fig.~\ref{Fig:zrel_hist}, but with 20\% of the KiDS Bright galaxies removed. {\it Upper panels:}   Galaxies from the cells with the mean relative difference between photometric and spectroscopic redshift from GAMA to be less [more] than  3.93\%, shown as a black [red] histogram. {\it Bottom panels:} Same as the upper panel, but the spectroscopic redshift is from the Espec sample.}
\label{Fig:zrel_hist20}
\end{figure}

\begin{figure}[h!]      
\centering
\includegraphics[width=0.49\textwidth]{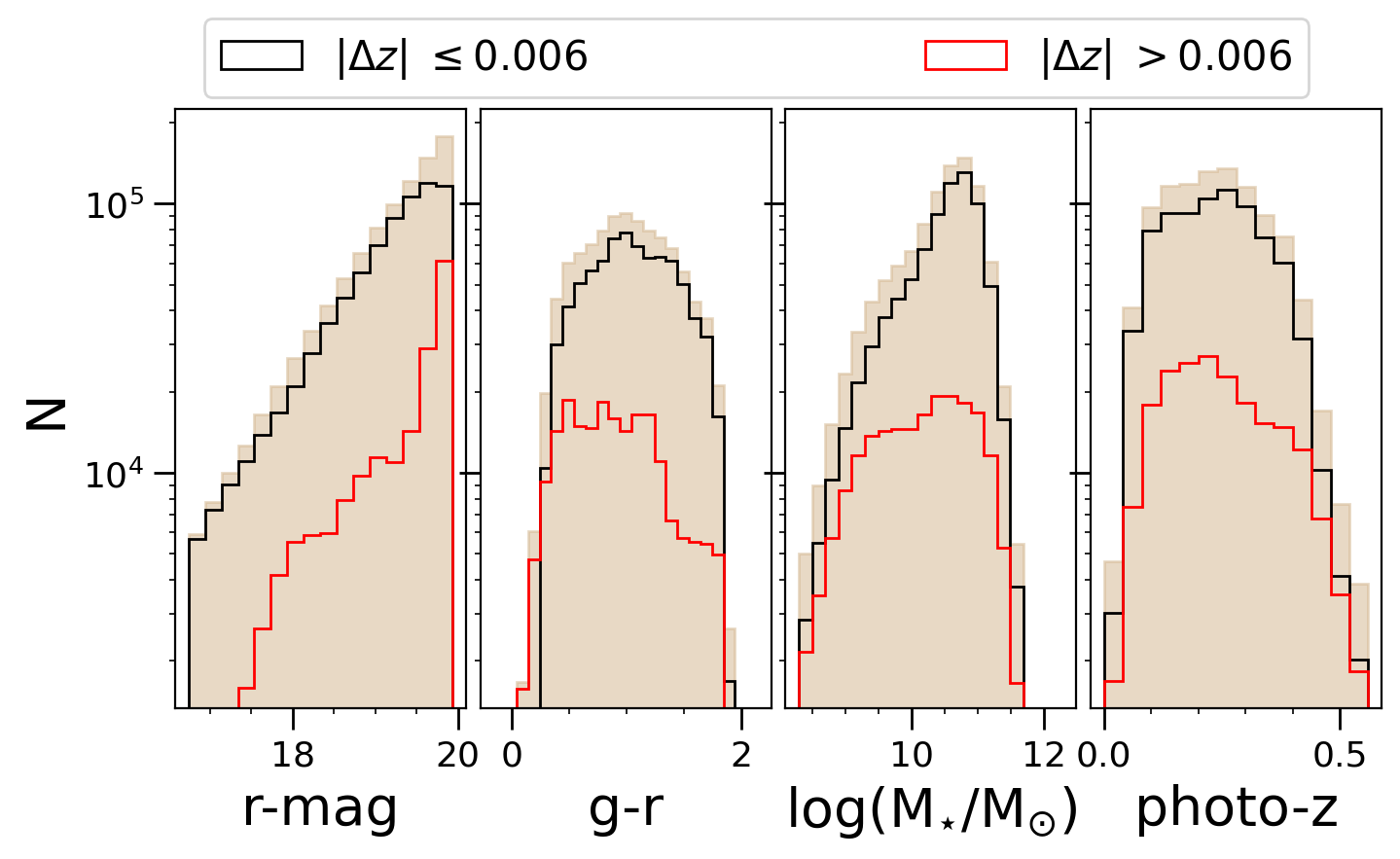}
\includegraphics[width=0.49\textwidth]{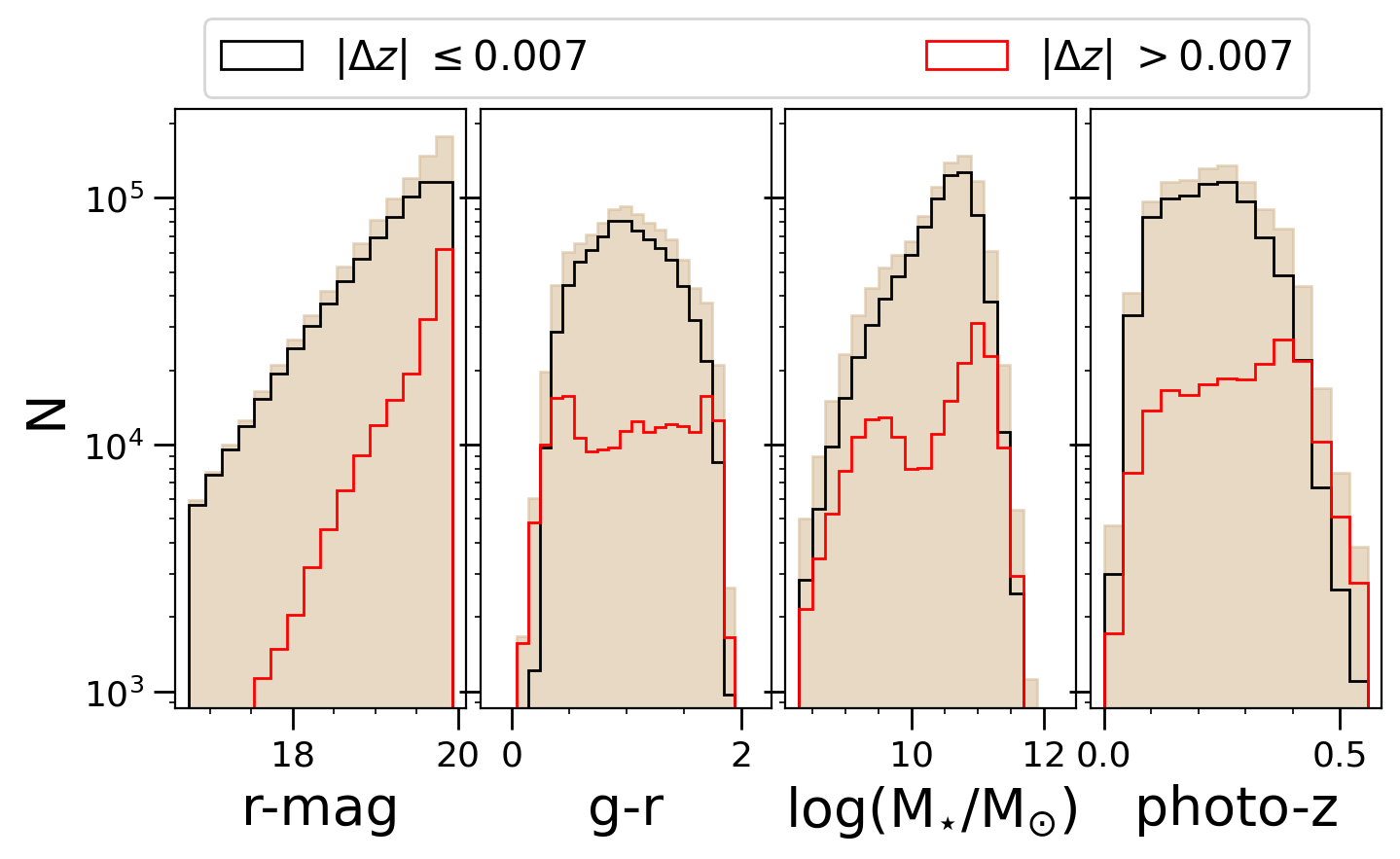}
\caption{Same as Fig.~\ref{Fig:zabs_hist}, but removing 20\% of galaxies with the mean absolute difference between photometric and spectroscopic redshift from GAMA to be less [more] than  0.006. The bottom panel shows the same but the mean absolute difference between photometric and spectroscopic redshift from Espec is less [more] than  0.007.}
\label{Fig:zabs_hist20}
\end{figure}

\begin{figure}[h!]
\centering
\includegraphics[width=0.49\textwidth]{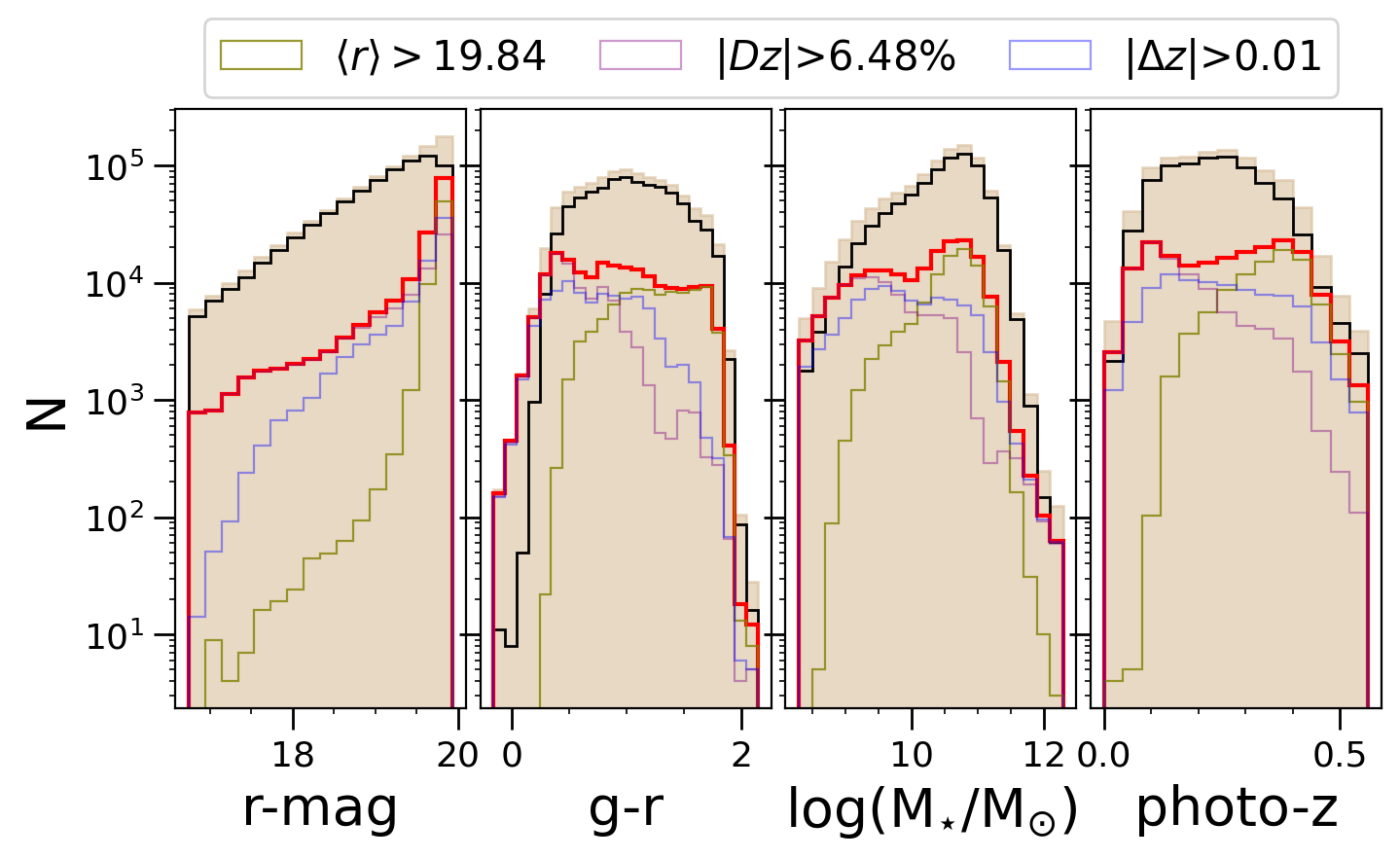}
\includegraphics[width=0.49\textwidth]{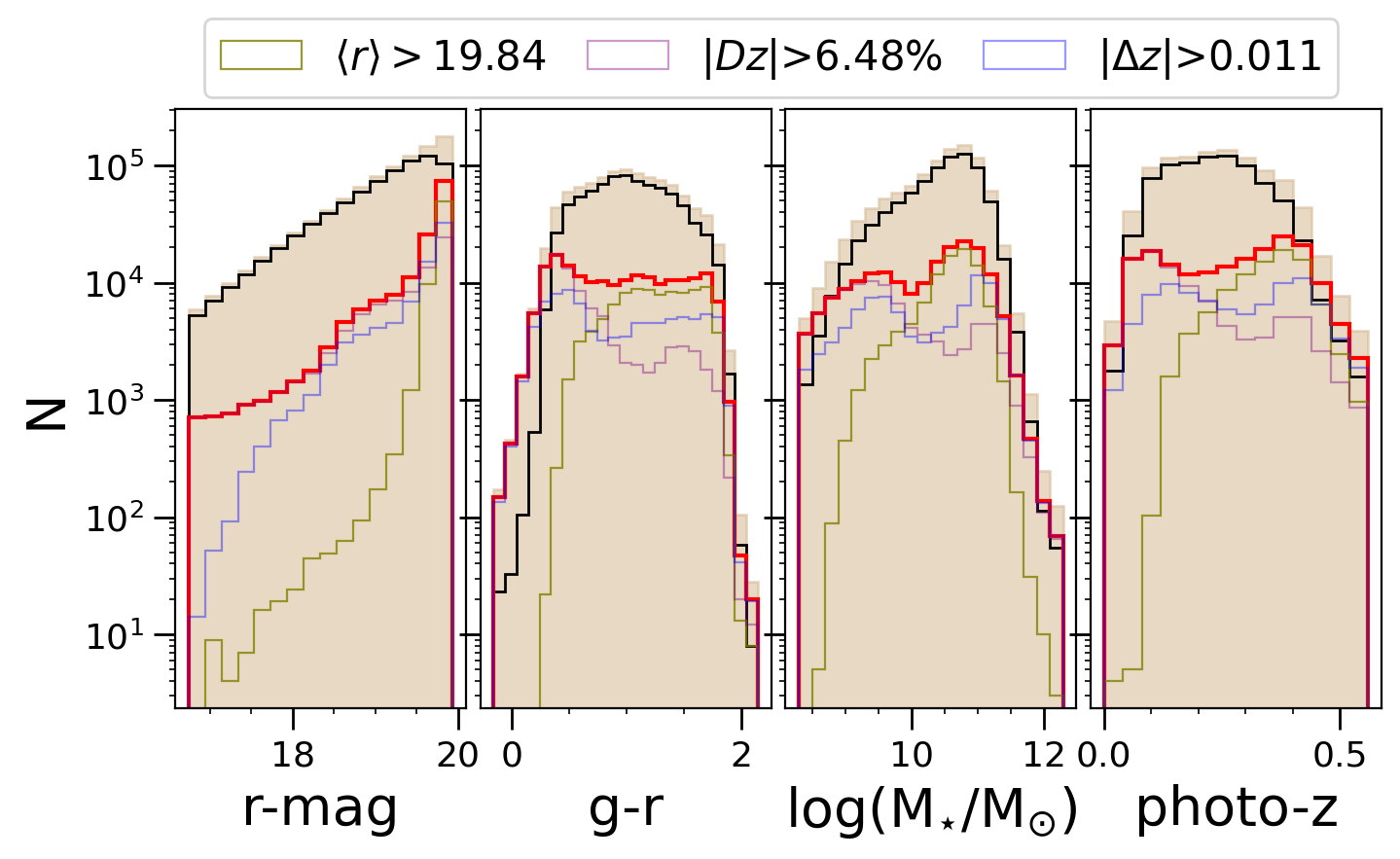}
\caption{Same as Fig.~\ref{Fig:joint} but instead of 10\% removal of galaxies, removing $\sim$19\% galaxies. {\it Upper panels:} The tan histograms show the distribution of KiDS galaxies versus r-magnitude, g-r color, stellar mass, and photometric redshifts. The red histogram shows the galaxies from the cells with either $\langle r \rangle >  19.84$ (green), $|Dz| > 6.48\%$ (purple), and based on $|\Delta z| > 0.01$ (blue), removing $\sim 19\%$ galaxies. The 80\% of the clean samples are shown in black. {\it Bottom panels:} Same as the upper panel, but the spectroscopic redshift is the mean from the Espec sample instead of GAMA in each SOM cell with $|Dz| > 6.48\%$ (purple), and based on $|\Delta z| > 0.011$, removing $\sim 19\%$ galaxies.}
\label{Fig:joint20}
\end{figure}

\begin{figure*}[h!]     
\centering
\includegraphics[width=\textwidth]{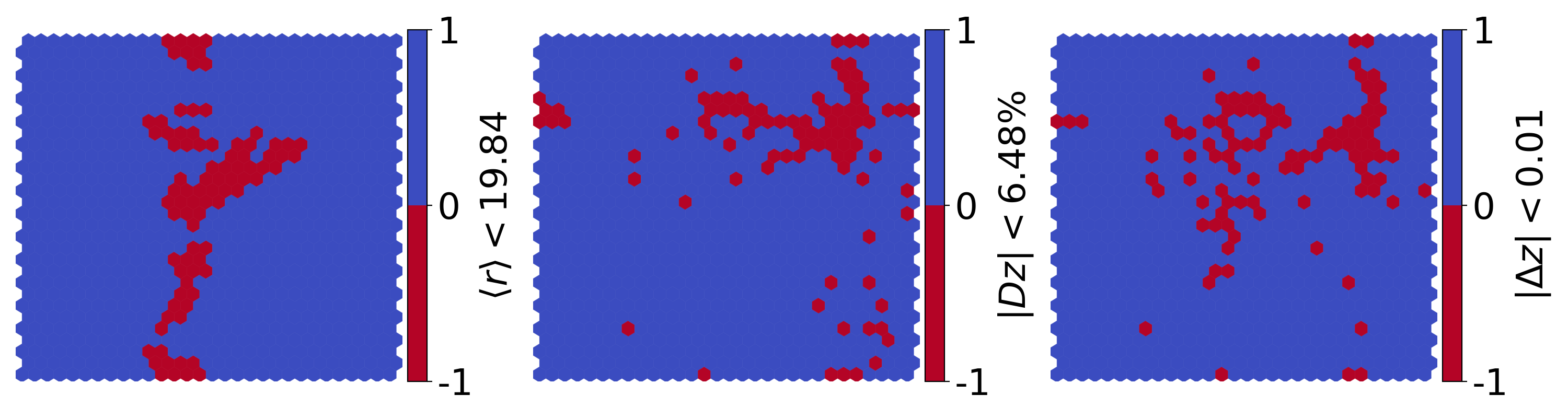}
\caption{Red SOM cells indicate the cells removed during the cut shown in Fig.~\ref{Fig:joint20}, removing 20\% of the KiDS-Bright galaxies. The significant overlap of cells in the $Dz$ and $\Delta z$ cuts is evident. The spec-$z$ are taken from GAMA.}
\label{Fig:somcells20}
\end{figure*}

\begin{figure*}[h!]     
\centering
\includegraphics[width=\textwidth]{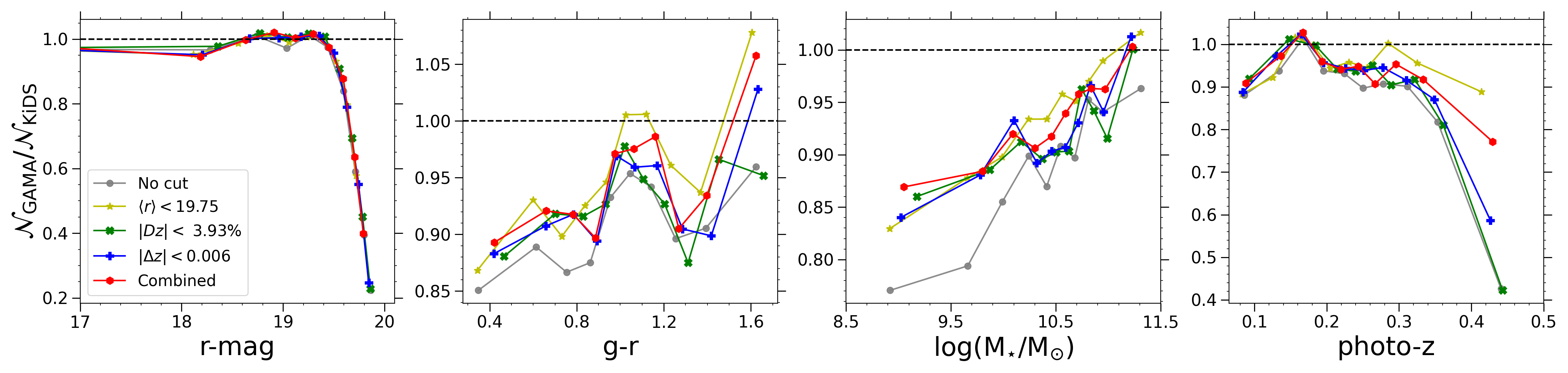}
\caption{Dependence of the GAMA completeness on selected properties of the KiDS-Bright sample, based on SOM projections. The completeness is expressed as the per-cell ratio of GAMA and KiDS-Bright surface density. The gray data points are the same as in Fig.~\ref{Fig:ratio_rzug}, illustrating binned medians for the full KiDS-Bright. The other colors show the completeness after applying cuts on the photometric sample, each employing SOM cell averages, namely:  (i) based on $\langle r \rangle <  19.75$   (yellow), (ii) based on relative difference $|Dz| < 3.93\%$   (green), (iii) based on $|\Delta z| < 0.006$   (blue), and finally (iv) a combined cut $\langle r \rangle < 19.84$, $|Dz| < 6.48\%$, $|\Delta z| < 0.01$ (red) removing $\sim20\%$ KiDS-Bright galaxy sample for each case.} 
\label{Fig:ratio_abscut20}
\end{figure*}

In Sect.~\ref{Sect:zcalib}, we present a method for cleaning up the KiDS-Bright sample for three main criteria, based on magnitude limit and photo-$z$ performance, and on their combination. In each case, we adjust the cuts in such a way as to remove roughly 10\% of the galaxies from the input dataset. Here we present the results of the same procedure, but for the case when we remove about 20\% objects from KiDS-Bright. The outcome is illustrated in Figs.~\ref{Fig:rcut_hist20}-\ref{Fig:joint20} and summarize in Table~\ref{Tab: photo-zstats20}, which are counterparts of those shown in Sect.~\ref{Sect:zcalib} for the $\sim10\%$ removal. We illustrate the SOM distribution of the removed cells in Fig.~\ref{Fig:somcells20} for a combination of 20\% galaxies removed. As can be seen, even such a more aggressive cut of keeping $\sim80\%$ of galaxies does not lead to an enormous improvement in photo-$z$ quality in terms of scatter (SMAD of $\delta z / (1+z)$), which is never reduced by more than 5\%  despite removing 1/5th of the KiDS-Bright galaxies. 
The completeness ratio shown in Fig.~\ref{Fig:ratio_abscut20} also shows the negligible improvement of the completeness ratio in stellar mass and photo-$z$ as compared to Fig.~\ref{Fig:ratio_rzug}.

We know that GAMA-Eq completeness, relative to KiDS-Bright, quickly declines for magnitudes fainter than $r\sim19.5$. Therefore, we also test the photo-$z$ statistics for two subsets $r>=19.5$ and $r<19.5$, as detailed in Table~\ref{Tab: photo-zstats20_r}. We find that, for the brighter subsample ($r < 19.5$), the mean redshift and deviations are generally lower, indicating better photometric redshift performance. The mean $\delta z/(1+z_\mathrm{spec})$ values are closer to zero, and both the standard deviations and SMAD are smaller compared to the fainter subsample.
For the fainter subsample ($r \geq 19.5$), the performance metrics show larger mean offsets and greater scatter, indicating challenges in achieving accurate photometric redshifts for these galaxies. Despite the selection cuts, the standard deviations and SMAD remain higher than those for the brighter subsample.

\begin{table*}[ht]
\begin{center}
\caption{Same as Table.~\ref{Tab: photo-zstats} but removing 20\% of the KiDS-Bright sample. 
\label{Tab: photo-zstats20}}
\begin{tabular}{lrrrrrr}
\hline
\centering { } {\textbf{Selections} } & { \textbf{Number of} } & \multicolumn{1}{c}{ \textbf{Mean} } & \multicolumn{1}{c}{ \textbf{Mean of} } & { \textbf{Mean of} } & \multicolumn{1}{c}{\textbf{St. dev.\ of}} & \multicolumn{1}{c}{\textbf{SMAD of}} \\

{} & { \textbf{galaxies} } & \multicolumn{1}{c}{ \textbf{redshift} } & \multicolumn{1}{c}{$\delta z = z_\mathrm{phot}-z_\mathrm{spec}$ } &  \multicolumn{1}{c}{$\delta z/(1+z_\mathrm{spec})$ } &  \multicolumn{1}{c}{$\delta z/(1+z_\mathrm{spec})$ } &  \multicolumn{1}{c}{$\delta z/(1+z_\mathrm{spec})$} \\ 

\hline    
No cut & $10.00\times10^5$ &  0.242 &   $4.7\times10^{-4}$ &  $9.1\times10^{-4}$ & 0.0232 & 0.0178 \\
$\langle r \rangle < 19.75$ &  $7.97\times10^5$ &  0.222 &   $4.4\times10^{-4}$ &  $8.7\times10^{-4}$ & 0.0227 & 0.0174 \\
$|Dz| <3.93\%$ (GAMA-based) &  $7.98\times10^5$ &  0.255 &  $-3.0\times10^{-4}$ &  $2.3\times10^{-4}$ & 0.0217 & 0.0174 \\
$|Dz| < 3.93\%$ (Espec-based) &  $8.01\times10^5$ &  0.250 &  $-2.5\times10^{-4}$ &  $2.5\times10^{-4}$ & 0.0214 & 0.0170 \\
$\Delta z < 0.006$ (GAMA-based) &  $7.99\times10^5$ &  0.239 &  $-2.8\times10^{-4}$ &  $2.5\times10^{-4}$ & 0.0217 & 0.0174 \\
$\Delta z< 0.007$ (Espec-based) &  $8.00\times10^5$ &  0.230 &  $-2.0\times10^{-4}$ &  $3.0\times10^{-4}$ & 0.0214 & 0.0171 \\
Combined (GAMA-based)$^1$ &  $8.07\times10^5$ &  0.237 &  $-1.6\times10^{-4}$ &  $3.3\times10^{-4}$ & 0.0215 & 0.0172 \\
Combined (Espec-based)$^2$ &  $8.11\times10^5$ &  0.234 &    $-1.0\times10^{-4}$ & $3.8\times10^{-4}$ & 0.0214 & 0.0172 \\
\hline
\end{tabular}
\begin{tablenotes}
$^1$ $\langle r \rangle < 19.84$ \&  $|Dz|$ $< 6.48\%$ \& $|\Delta z| < 0.010$. \\
$^2$ $\langle r \rangle < 19.84$ \&  $|Dz|$ $< 6.48\%$ \& $|\Delta z| < 0.011$.    
\end{tablenotes}
\end{center}
\end{table*}

\begin{table*}
\begin{center}
\caption{Same as Table.~\ref{Tab: photo-zstats20} but sample divided in two $r$ bins $r<19.5$ [$r\geq 19.5$]. 
\label{Tab: photo-zstats20_r}}
\begin{tabular}{lrrrrrr}
\hline
\centering { } {\textbf{Selections} } & { \textbf{Number of} } & \multicolumn{1}{c}{ \textbf{Mean} } & \multicolumn{1}{c}{ \textbf{Mean of} } & { \textbf{Mean of} } & \multicolumn{1}{c}{\textbf{St. dev.\ of}} & \multicolumn{1}{c}{\textbf{SMAD of}} \\

{} & { \textbf{galaxies} } & \multicolumn{1}{c}{ \textbf{redshift} } & \multicolumn{1}{c}{$\delta z = z_\mathrm{phot}-z_\mathrm{spec}$} &  \multicolumn{1}{c}{$\delta z/(1+z_\mathrm{spec})$ } &  \multicolumn{1}{c}{$\delta z/(1+z_\mathrm{spec})$ } &  \multicolumn{1}{c}{$\delta z/(1+z_\mathrm{spec})$} \\ 
{} & { \textbf{($\times10^5$)} } & \multicolumn{1}{c}{ \textbf{} } & \multicolumn{1}{c}{($\times10^{-4}$)} &  \multicolumn{1}{c}{($\times10^{-4}$)} &  \multicolumn{1}{c}{} &  \multicolumn{1}{c}{} \\ 

\hline   
No cut & $5.87~[4.13]$ &  0.209 [0.288] & $1.7~[13.3]$ &  $6.0~[17.6]$ & 0.0216 [0.0273] & 0.0166 [0.0216] \\
$\langle$ r $\rangle < $19.75 & $5.81~[2.16]$ &  0.208 [0.260] & $1.5~[16.1]$ &  $5.8~[19.9]$ & 0.0214 [0.0270] & 0.0166 [0.0211]\\
$|Dz|$ $<$3.93\% (GAMA-based) & $4.81~[3.17]$ &  0.221 [0.307] & $-6.8~[7.8]$ & $-1.4~[12.7]$ & 0.0202 [0.0254] & 0.0163 [0.0208]\\
$|Dz|$ $< $3.93\% (Espec-based) & $4.87~[3.15]$ &  0.221 [0.296] & $-6.1~[8.0]$ & $-0.9~[12.6]$ & 0.0198 [0.0252] & 0.0160 [0.0206]\\
$\Delta z$ $< $0.006 (GAMA-based) & $5.08~[2.91]$ &  0.210 [0.290] & $-5.4~[5.2]$ & $-0.1~[10.5]$ & 0.0204 [0.0252] & 0.0165 [0.0208]\\
$\Delta z$ $ < $0.007 (Espec-based) & $5.15~[2.85]$ &  0.206 [0.274] & $-3.8~[3.9]$ &   $1.0~[9.3]$ & 0.0201 [0.0250] & 0.0162 [0.0206]\\
Combined (GAMA-based)$^1$ & $5.42~[2.65]$ &  0.214 [0.283] & $-4.6~[8.4]$ &  $0.4~[12.9]$ & 0.0202 [0.0251] & 0.0163 [0.0207]\\
Combined (Espec-based)$^2$ & $5.39~[2.72]$ &  0.214 [0.274] & $-4.0~[8.7]$ &  $0.9~[13.2]$ & 0.0201 [0.0252] & 0.0162 [0.0208]\\
\hline
\end{tabular}
\begin{tablenotes}
$^1$ $\langle r\rangle < 19.84$ \&  $|Dz|$ $< 6.48\%$ \& $|\Delta z| < 0.010$. \\
$^2$ $\langle r\rangle < 19.84$ \&  $|Dz|$ $< 6.48\%$ \& $|\Delta z| < 0.011$.    
\end{tablenotes}
\end{center}
\end{table*}

\section{Dependence of photo-$z$ quality on the percentage of removed galaxies}
\label{app:percentages}

\begin{figure*}     
\centering
\includegraphics[width=0.4\textwidth]{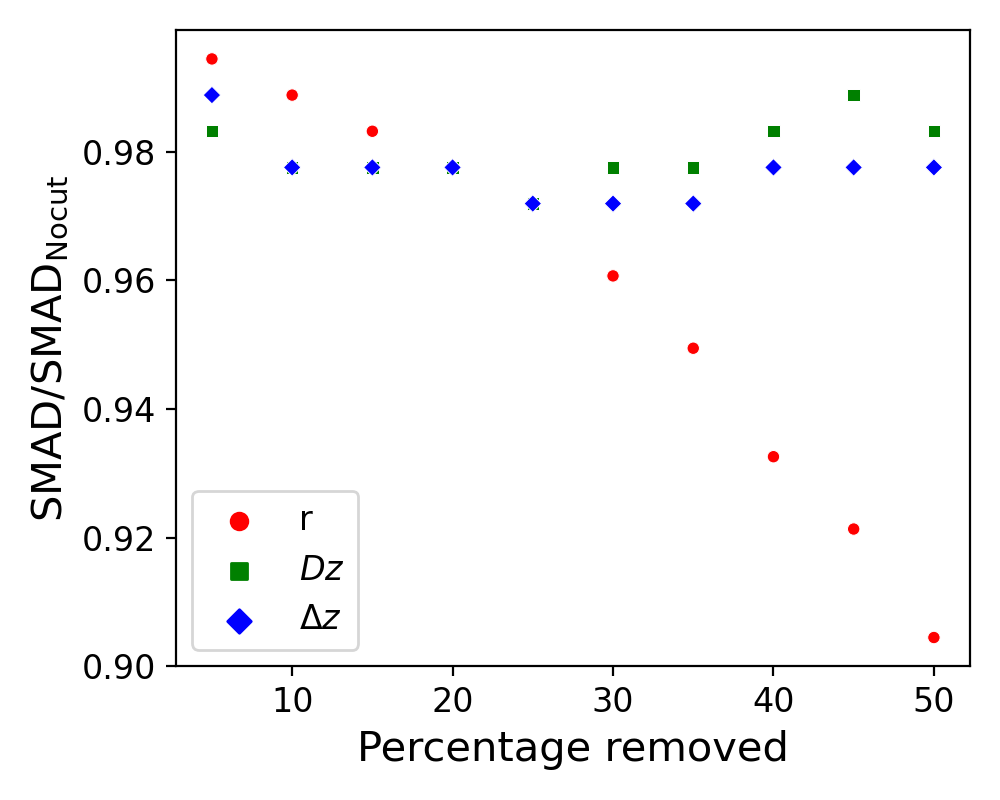}
\includegraphics[width=0.4\textwidth]{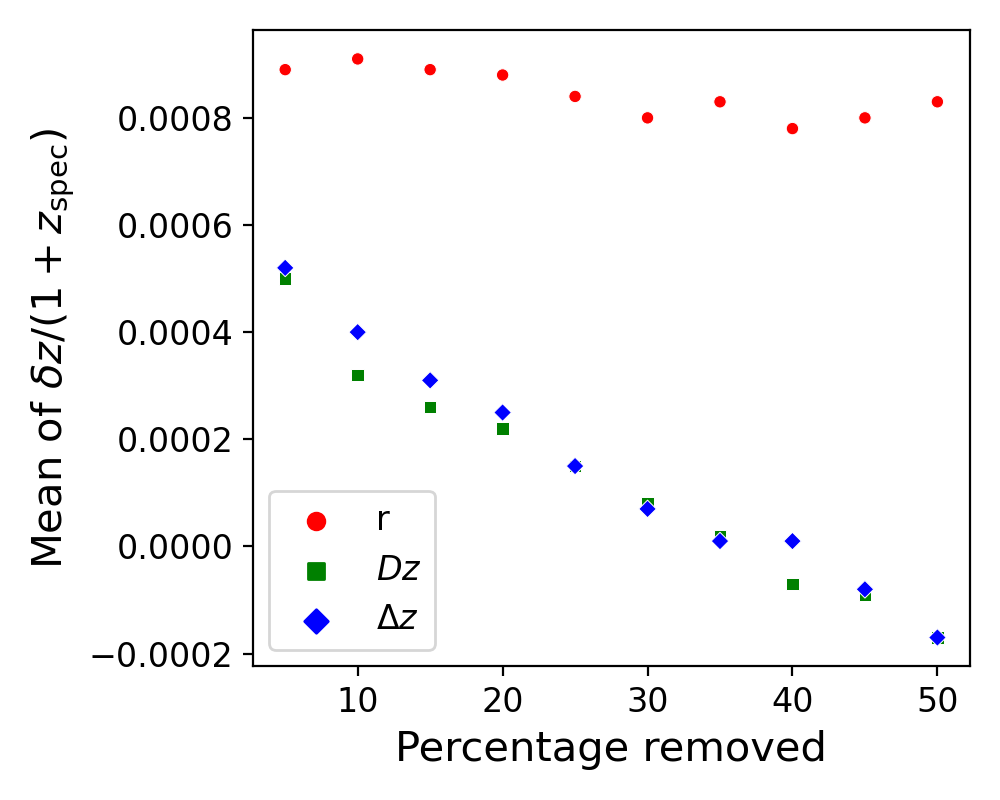}
\caption{Relative change in the SMAD of $\delta z / (1+z)$ of the KiDS-Bright sample as a function of the fraction of removed galaxies (left),
after applying cuts (based on GAMA) on the photometric dataset, each employing SOM cell averages, based on $\langle r\rangle$  (blue-circle), relative difference $|Dz|$ (orange-square) and absolute $|\Delta z|$ (green-diamond). \textit{Right panel}: Variation in the mean photo-$z$ residual for the same setups as in the left panel.} 
\label{Fig:smad_bias_frac}
\end{figure*}

Here, we generalize the possible sample cleanup and check how many galaxies we should remove from KiDS-Bright to obtain the desired reduction in photo-$z$ bias or scatter. This is summarized in Fig.~\ref{Fig:smad_bias_frac}, where we show the SMAD and mean photo-$z$ residual (both based on GAMA) for various percentages of removed galaxies, starting from 5\% up to as much as 50\%, which ends up reducing the dataset by half. As the left panel indicates, we find that SMAD values are consistently lowered if we remove the faint end, and this reduction is up to 10\% but only when the faintest half of the sample is removed ($\langle r \rangle >19.36$ mag). On the other hand, cuts $\Delta z$ or $Dz$ hardly affect the SMAD which changes by less than 3\% irrespective of the culled fraction. The picture is reversed if we consider the change in mean residuals. In the right panel of Fig.~\ref{Fig:smad_bias_frac} we can see that $\langle \Delta z \rangle$ stays at the level of $\sim8\times10^{-4}$ no matter how many faint galaxies are removed. On the other hand, this photo-$z$ residual goes down roughly linearly with more aggressive cuts in SOM-based $Dz$ and $\Delta z$. In fact, at around 40 - 45\% of removed galaxies, it crosses the zero line and becomes negative. 

\end{document}